\documentclass[twocolumn,showpacs,prX,nofootinbib,nobibnotes]{revtex4}

\usepackage{graphicx}
\usepackage{dcolumn}
\usepackage{bm}

\usepackage{anysize}
\usepackage{graphicx}
\marginsize{1.9 cm}{1.4 cm}{0.5 cm}{3.0 cm}

\usepackage{graphicx}
\usepackage{dcolumn}
\usepackage{bm}
\usepackage{amsmath}
\usepackage{amssymb}
\usepackage{mathrsfs}
\usepackage[latin1]{inputenc}
\usepackage{amsmath}

\newcommand{\p}{\partial}
\newcommand{\dd}{{\rm d}}
\newcommand{\e}{\varepsilon}
\newcommand{\sd}{Schr\"{o}dinger }

\newcommand{\U}{\mathcal{U}}

\newcommand{\tr}{{\rm Tr}}

\newtheorem{theorem}{Theorem}
\newtheorem{lemma}{Lemma}

\begin{document}


\title{Quantum Pareto Optimal Control}

\author{Raj Chakrabarti}\email{rajchak@princeton.edu}
\affiliation{Department of Chemistry, Princeton University,
Princeton, New Jersey 08544, USA}

\author{Rebing Wu}
\affiliation{Department of Chemistry, Princeton University,
Princeton, New Jersey 08544, USA}

\author{Herschel Rabitz}
\affiliation{Department of Chemistry, Princeton University,
Princeton, New Jersey 08544, USA}

\date{20 May 2008}

\begin{abstract}

We describe algorithms, and experimental strategies, for the Pareto optimal control problem of simultaneously driving
an arbitrary number of quantum observable expectation values to their respective extrema.
Conventional quantum optimal control strategies are less effective at sampling points on the Pareto frontier of
multiobservable control landscapes than they are at locating optimal solutions to single observable control problems.
The present algorithms facilitate multiobservable optimization by following direct paths to the Pareto front, and are capable of continuously tracing the front once it is found to explore families of viable solutions.
The numerical and experimental methodologies introduced are also applicable to other problems that require the
simultaneous control of large numbers of observables, such as quantum optimal mixed state preparation.
\end{abstract}


\maketitle


\section{Introduction}\label{intro}

The methodology of quantum optimal control has been applied extensively to problems requiring the maximization of the
expectation values of single quantum observables. Recently, an important new class of
quantum control problems has surfaced, wherein the aim is the simultaneous maximization of the expectation values of
multiple quantum observables, or more generally, control over the full quantum state as encoded in the density matrix.
Whereas various algorithms for the estimation of the density matrix have been reported in the literature
\cite{Banaszek1999}, comparatively little attention has been devoted to the optimal control of the density matrix or
multiobservables \cite{Wolf2007,Weber2007,RajWu2008}.
Such methodologies are important in diverse applications including the control of product selectivity in coherently
driven chemical reactions \cite{Weber2007}, the dynamical discrimination of like molecules \cite{Li2002,Li2005}, and
the precise preparation of tailored mixed states.

In a recent work \cite{RajWu2008}, we reported an experimentally implementable control methodology - quantum
multiobservable tracking control (MOTC) - that can be used to simultaneously drive multiple observables to desired
target expectation values, based on control landscape gradient information.  It was shown that, due to the fact that the critical manifolds of quantum control landscapes \cite{Raj2007} have measure zero in the search domain, tracking
control algorithms are typically unobstructed when following paths to arbitrary targets in multiobservable space.
Therefore, these algorithms can facilitate multiobservable control by following direct paths to the target expectation
values, and are typically more efficient than algorithms based on optimization of a control cost functional. The MOTC
method falls into the general category of continuation algorithms for multiobjective optimization
\cite{Hillermeier2001}. These were introduced several years ago as alternatives to stochastic multiobjective
optimization algorithms, which do not exploit landscape gradient information.

In this paper, we extend the theory of MOTC to complex multiobjective optimization problems such as multiobservable maximization, and propose experimental techniques for the implementation of MOTC in such complex multiobservable control scenarios. Multiobjective maximization typically seeks to identify non-dominated (rather than completely optimal) solutions, which lie on the so-called Pareto frontier \cite{Statnikov1990}.  Conventional multiobjective control algorithms require substantial search effort in order to adequately sample the Pareto frontier and locate solutions that strike the desired balance between the various objectives. Direct application of optimal control (OC) \cite{Levis2001} or MOTC algorithms is also inefficient in other applications, such as quantum state preparation, that require control of a large number $m$ of observables.

When naively applied, both OC and MOTC fail to exploit the simple underlying geometry of quantum
control landscapes \cite{Raj2007}, by convoluting a) the dynamic \sd map $\varepsilon(t) \mapsto U(T)$ between
time-dependent control fields and associated unitary propagators at the prescribed final time $T$ with b) the kinematic map $U(T) \mapsto \Phi(U(T))$ between unitary propagators and associated observable expectation values.
In multiobservable Pareto optimal control, the Pareto frontier on the domain $\U(N)$ has a simple geometric
structure, which can be exploited to efficiently sample the corresponding frontier on the domain $\e(t)$.
Here, we develop strategies that combine the application of MOTC with quantum state estimation and kinematic optimization on $\U(N)$ for the purpose of solving Pareto optimal control and other large $m$ observable control problems.

The paper is organized as follows. In Section II, we provide necessary preliminaries on Pareto optimal control and
quantum multiobservable control cost functionals. Section III presents analytical results on the distribution of
Pareto optima in quantum multiobservable maximization problems. In Section IV,
we describe how MOTC can be used to locate corresponding points on the dynamic Pareto front and to subsequently explore families of fields within the front that minimize auxiliary, experimentally relevant costs. Section V proposes measurement strategies aimed at improving the experimental efficiency of MOTC.  In Section VI, we numerically implement Pareto optimal tracking control and illustrate the use of efficient measurement strategies for difficult problems with large numbers of observables $m$. We also examine the impact of prior state preparation and the advantages of using MOTC versus scalar cost function optimization for Pareto front sampling. Finally, in Section VII, we discuss the implications of these results for Pareto optimal control experiments.

\section{Pareto optimization of quantum control cost functionals}\label{gradientflows}

The class of quantum optimal control problems we examine in this paper in can be expressed
\cite{Raj2007}:
\begin{equation}\label{OCT}
\max_{\e(t)}~~\Phi(U(T)).
\end{equation}
$U(T)$ is an implicit functional of $\e(t)$ via the \sd
equation for the unitary propagator
$$\frac{\dd U(t)}{\dd t}=-
\frac{\imath}{\hbar}H(\e(t))U(t), ~~~U(0)=I_N,$$ where $H$ is the
total Hamiltonian, and $\e (t)$ is the time-dependent control field.
Solutions to the optimal control problem correspond to $\frac{\delta
\Phi}{\delta \e(t)} = 0$ for all $t\in[0,T]$.
In quantum single observable control, $\Phi$ is typically taken to be the
expectation value of an observable of the system:
\begin{equation}\label{obs}
\Phi(U(T)) = \tr(U(T){\rho(0)}U^{\dag}(T)\Theta),
\end{equation}
where $\rho(0)$ is the density matrix of the system at time $t=0$ and
$\Theta$ is a Hermitian observable operator whose expectation
value we seek to maximize \cite{Mike2006a}. Recent work \cite{Raj2007} has demonstrated
that the landscape for optimization of (\ref{obs}) is devoid of local extrema.

Although the goal of a multiobjective optimization  
problem may be to maximize the expectation values of all
observables, i.e.,
\begin{equation*}
\max_{\e(t)} ~\vec{\Phi}(U(T)) =
\{\Phi_1(U(T),\cdots,\Phi_m(U(T))\},
\end{equation*}
in many cases the $\Phi_{k}$ cannot be simultaneously maximized.
Thus, the scalar concept of optimality must be replaced by the
notion of Pareto optimality \cite{Statnikov1990}. A control field
$\e^*(t)$ is said to be \textit{strongly Pareto optimal} if all other fields 
$\e(t)$ have a lower value for at least one of the objective
functions $\Phi_k(\cdot)$, or else have the same value for all
objectives. $\e^*(t)$ is \textit{weakly Pareto optimal} if there does not exist
another field $\e(t)$ such that $\Phi_k(\e(t)) > \Phi_k(\e^*(t))$ for all $k$. Analogous definitions
hold for strongly and weakly Pareto optimal unitary propagators $U^*(T)$, which
we refer to as kinematic Pareto optima in order to distinguish them from the aforementioned dynamic Pareto optima. Fig. 1 provides examples of strong and weak kinematic Pareto optima.
We denote the set of strong  (kinematic, dynamic) Pareto optima by $\mathcal{P}^{(U,\e)}_s$, and the set of weak Pareto
optima by $\mathcal{P}^{(U,\e)}_w$.
We will use the term \textit{Pareto frontier} (or Pareto front) to refer to $\mathcal{P}^{(U,\e)}_s$ or
$\mathcal{P}^{(U,\e)}_w$.
Our primary interest in this paper is the development of algorithms for sampling weak Pareto optima,
which are much more numerous in quantum control problems than are strong Pareto optima.
Note that it is possible to define the notion of Pareto optimality for more general cost functions; in particular,
selected observable expectation values may be minimized through the replacements $\Phi_k(\cdot) \rightarrow
-\Phi_k(\cdot)$.

A natural scalar objective function $\Phi$ for multiobservable control is a
positively weighted convex sum of the individual observable objectives,
i.e.,
\begin{equation}\label{multi}
\Phi_{M}(U) = \sum_{k=1}^m \alpha_k \Phi_{k}(U), \quad \alpha_k > 0,
\end{equation}
where $\Phi_{k}= \tr(U(T)\rho(0)U^{\dag}(T)\Theta_k), \quad
k=1,2,\cdots,m$.
Within the electric dipole
formulation, where the Hamiltonian assumes the form
\begin{equation}\label{ham}
H(t) = H_0 - \mu \cdot \e(t)
\end{equation} with internal Hamiltonian $H_0$ and electric dipole operator $\mu$, the gradient of $\Phi_M$ is
\cite{HoRab2007a}:
\begin{equation}\label{gradmult}
\frac{\delta \Phi_M}{\delta \e(t)} = -\frac{\imath}{\hbar}\sum_k
\alpha_k \tr\{\left[\Theta_k(T),\mu(t)\right]\rho(0)\},
\end{equation}
where $\mu(t)= U^{\dag}(t)\mu U(t)$.
One approach to quantum Pareto optimal control is to employ gradient flow algorithms of the form
\begin{equation}\label{Eflow}  %
\frac{\p\e_s(t)}{\p s}= \gamma \frac {\delta \Phi_{M}}{\delta \e_s(t)},
\end{equation}
where $\gamma$ is an arbitrary positive constant and $s$ is a
continuous variable parameterizing the algorithmic search
trajectory, to maximize an objective function or the form
(\ref{multi}). In order to sample the Pareto front, various settings
of the convex weights in (\ref{multi}) are typically sampled in
independent optimizations \cite{Lin1976}. Note that the objective
function $\Phi_M$ is just the expectation value of a single quantum
observable $\Theta_M = \sum_{k=1}^m \alpha_k \Theta_k$. The absence
of local traps \cite{Raj2007}  in control landscapes for the single
observable objective function (\ref{obs}) thus indicates that the
multiobservable gradient flow will also reach its global optimum
\footnote{It is possible to construct multiobservable objective
functions that have a more general nonlinear form, such as those
that have been used in optimal dynamical discrimination
\cite{Li2005}. These objective functions may possess diverse
landscape topologies. Nonetheless, the most commonly employed
multiobjective functions are devoid of traps.}. However, it is not
immediately clear whether the multiobservable gradient flow
(\ref{Eflow}) converges to points on the Pareto front, since the
global optima of ($\ref{obs}$) for $\Theta=\Theta_k$ and
($\ref{multi}$) do not necessarily intersect. The same question
arises for stochastic optimization of (\ref{multi}).   We now
examine the conditions under which (\ref{Eflow}) will converge to a
weak Pareto optimum.

\section{Identification of Pareto optima}\label{kinpareto}

In order to understand the structure of the set $\mathcal{P}^{U}_w$ of weakly Pareto optimal points, it is necessary
to first characterize the global optima of the individual objectives.
The condition for a unitary propagator $U$ to be a critical point of objective function (\ref{obs}) is
\cite{WuMike2008}:
\begin{equation}\label{critcond}
[\Theta_k,U\rho(0)U^{\dag}]=0.
\end{equation}
Let $\rho(0)$ and $\Theta_k$ be diagonalized according to $\tilde \rho =
R^{\dag}\rho(0) R$, $\tilde \Theta_k = S_k^{\dag}\Theta_k S_k$, such that
\begin{eqnarray}\label{diagonal}
\tilde \rho &=& diag\{\lambda_{(1)},\cdots,\lambda_{(1)};\cdots;\lambda_{(r)},\cdots,\lambda_{(r)}\}\\
\tilde \Theta_k &=& diag\{\gamma_{k(1)},\cdots,\gamma_{k(1)};\cdots;\gamma_{k(s_k)},\cdots,\gamma_{k(s_k)}\},
\end{eqnarray}
where the $\lambda_{(i)},~\gamma_{k(i)}$ are distinct eigenvalues of $\rho(0), \Theta_k$ with multiplicities
$n_1,\cdots,n_r$ and $m_{k1},\cdots,m_{ks_k}$, respectively. Further, set $\tilde U \equiv R^{\dag}US_k$. We choose $R$
such that $\lambda_{(1)}<\cdots<\lambda_{(r)}$ and $S_k$ such that $\gamma_{k(1)}<\cdots<\gamma_{k(s_k)}$. A diagonal matrix of eigenvalues satisfying this condition is said to
be ``arranged in increasing order".

It was shown in \cite{WuMike2008} that the critical submanifolds
$\mathcal{M}_k^i$ that satisfy equation (\ref{critcond}) for each
objective $\Phi_k$ may be expressed in terms of $\tilde U$ as the
double cosets
\begin{equation}\label{quotient}
\mathcal{M}_k^i=\U(\textbf{n}) \pi^i \U(\textbf{m}_k),
\end{equation}
where $\U(\textbf{m}_k)$ is the product group $\U(m_{k1}) \times
\cdots \times \U(m_{ks_k})$, each $\U(m_{kl})$ corresponding to an
eigenvalue of $\Theta_k$ with $m_{kl}$-fold degeneracy (respectively
$\U(\textbf{n}),\U(n_{1}) \times \cdots \times
\U(n_{r}),~\rho(0)$,$~n_l$), and $\pi^i$ denotes a permutation
matrix on $N$ indices. We denote by $diag\{\gamma_{k1}^{\pi},\cdots,\gamma_{kN}^{\pi}\}=\pi^{\dag}\tilde \Theta_k\pi$
the diagonal matrix of (possibly degenerate) eigenvalues resulting from application of a permutation $\pi$ to $\tilde
\Theta_k$.

If two permutation matrices $\pi^1, \pi^2$
satisfy the relation
\begin{equation}\label{equiv}
\pi^2 = U \pi^1 V, \quad (U,V) \in  \U(\mathbf{n}) \times \U(\mathbf{m}_k),
\end{equation}
then they are associated with the same critical submanifold, and we
write $\pi^1 \sim \pi^2$. By this equivalence relation, we can
partition the permutation group $\mathcal{D}(N)$ into nonintersecting
subsets:
\begin{equation}\label{partition}
\mathcal{P}(N)=\mathcal{D}_k^1\cup\cdots\cup\mathcal{D}_{k}^{d_k},
\end{equation}
where each subset is an equivalence class with respect to $\sim$ and
uniquely corresponds to a critical submanifold; $d_k$ is the number
of critical submanifolds. For simplicity, we always let
$\mathcal{D}_k^{\max}=\mathcal{D}_k^1$ correspond to the optimum
manifold $\mathcal{M}_k^{\max}$, whose elements arrange the
eigenvalues of $\tilde\Theta_k$ in increasing order.

Analytical results concerning the distribution of Pareto optima are most readily derived in the case where the
observables $\{\Theta_k\}$ commute. In this case,
they can be simultaneously diagonalized by a single unitary transformation $S$.
Because the partition (\ref{partition}) for each $k$ is complete,
each $\mathcal{D}_k^{i}$ must overlap with at least one $\mathcal{D}_{k'}^{j}$.
Hence, the critical manifolds $\bigcup_i \mathcal{M}_k^i$ and $\bigcup_j \mathcal{M}_{k'}^j$ of $\Phi_k$ and $\Phi_{k'}$, respectively, must intersect. The critical \textit{sub}manifolds $\mathcal{M}_k^i$ and $\mathcal{M}_{k'}^j$ intersect if the corresponding $\mathcal{D}_k^i\cap\mathcal{D}_{k'}^j\neq \emptyset$. 

Based on the above result, we can specify the relationship between
the maximum of the function $\Phi_M(U)$ and the critical
submanifolds of the single observable cost functions $\Phi_k(U)$ in
the case that the $\{\Theta_k\}$ commute. Let $\Theta_M \equiv
\sum_{k=1}^m \alpha_k \Theta_k$, let
$\mathcal{P}(N)=\mathcal{D}_M^1\cup\cdots\cup\mathcal{D}_M^{d_M}$ be
the corresponding partition of permutation group, and let $S$ denote the
unitary transformation that arranges $\tilde \Theta_M$ in increasing order. Then we have the
following lemma pertaining to the conditions for convergence of
gradient flow (\ref{Eflow}) to a Pareto optimum.

\begin{lemma}
Let $\{\Theta_k\}$ be a set of mutually commuting observables. The
gradient flow (\ref{Eflow}) may (depending on initial conditions)
converge arbitrarily close to a weak Pareto optimum if for at least one $k$,
$\mathcal{D}_k^{\max}\cap \mathcal{D}_M^{\max}\neq\emptyset$. The flow is guaranteed to converge to a weak Pareto optimum if for each $~\U(m_{Mi})$ we have $S_k^{\dag}S ~\U(m_{Mi}) ~S^{\dag} S_k \subseteq \U(m_{kj})$ for some $j$, and $\mathcal{D}_M^{\max} \subseteq \mathcal{D}_k^{\max}$. The flow may converge arbitrarily close to a strong Pareto optimum if $\mathcal{D}_k^{\max}\cap \mathcal{D}_M^{\max}\neq\emptyset$ for all $k$.
For each $\Theta_k$, the flow may converge arbitrarily close to the critical submanifold $\mathcal{M}_k^{i}$ of $\Phi_k$ if $\mathcal{D}_k^{i}\cap \mathcal{D}_M^{\max}\neq\emptyset$.
\end{lemma}
The proof for guaranteed convergence to $\mathcal{P}_w^{(U,\e)}$ is presented in Appendix A. The remaining claims follow directly from the arguments above.

Now consider the problem of solving for the coefficients  $\{\alpha_k\}$  such that the gradient flow is capable of
converging to an arbitrarily specified combination of $l \leq m$ critical submanifolds of the respective observable
operators, i.e.,
\begin{equation}\label{condset}
\Phi_k(U) = \chi_k^{i_k},\quad \chi_k^{i_k} \in \mathcal{C}_k, \quad 1 \leq k \leq l,
\end{equation}
where $\chi_k^{i_k}$ denotes the $i_k$-th critical value in the set
$\mathcal{C}_k$ of critical values of $\Phi_k(U)$.
Let us write $\tilde \Theta_M = S^{\dag}\Theta_M S
=diag\{\sum_k\alpha_k\gamma_{k1}^{\pi_k},\cdots,\sum_k\alpha_k\gamma_{kN}^{\pi_k}\}$, where $\pi_k$ is the permutation
matrix that reorders the diagonal elements of $\tilde \Theta_k = S_k^{\dag}\Theta_k S_k$ so that they are arranged in
the order induced by $S$, i.e., $S = \pi_k^{\dag} S_k \pi_k$.
Provided that we have
\begin{equation}\label{picond}
\mathcal{D}_1^{i_1}\cap \cdots\cap\mathcal{D}_l^{i_l}\neq\emptyset,
\end{equation}
then for each permutation matrix $\pi\in\mathcal{D}_1^{i_1}\cap
\cdots\cap\mathcal{D}_l^{i_l}$, there is an independent system of
$N!-1$ inequalities:
\begin{equation}\label{inequalities}
\sum_{j=1}^N \sum_{k=1}^m \alpha_k\lambda_j \left(\gamma_{kj}^{\pi} - \gamma_{kj}^{\pi'}\right) \geq 0, ~~ \forall
\pi'\neq \pi. 
\end{equation}
It is possible to satisfy conditions (\ref{condset}) at the global optimum of the objective function (\ref{multi}) if a solution to this set of inequalities exists for at least one such $\pi$.
We therefore have the following theorem.

\begin{theorem}
Consider a quantum Pareto optimization problem of the form
(\ref{condset}) involving mutually commuting observables
$\{\Theta_k\}$. \begin{enumerate} \item It is possible to design an
objective function (\ref{multi}) whose gradient flow may (depending
on initial conditions) converge arbitrarily close to a weak Pareto
optimum corresponding to $\chi_k^{i_k}=\chi_k^{\max}$ while
arranging the expectation values of observables $\Theta_{k'},~ k'
\neq k$ in the order designated by equation (\ref{condset}), if a
solution to the inequalities (\ref{inequalities}) exists for at
least one $\pi \in \mathcal{D}_1^{i_1}\cap
\cdots\cap\mathcal{D}_l^{i_l}.$ \item Similarly, it is possible to
minimize the expectation value of observable $\Theta_k$ if a
solution exists for $\chi_k^{i_k}=\chi_k^{\min}$. \item If a
solution exists for the choice $\{\chi_k^{i_k}\} =
\{\chi_k^{\max}\}$, then it is possible to design an objective
function (\ref{multi}) whose gradient flow converges arbitrarily
close to a strong Pareto optimum.\end{enumerate}
\end{theorem}
The results of Theorem 1 pertaining to convergence to Pareto optima hold also for stochastic search
algorithms that optimize a conventional multiobjective cost function of the form (\ref{multi}).

The relative likelihoods of the gradient flow (\ref{Eflow}) converging to various Pareto optima can be
qualitatively assessed by comparing the dimension of $\mathcal{M}_M^{\max}$ to the dimension of the intersection $\mathcal{I}_{M,k}^{\max,\max}=\mathcal{M}_M^{\max} \cap \mathcal{M}_k^{\max}$ for each observable $k$ whose $\chi_k^{i_k}=\chi_k^{\max}$ in equation (\ref{condset}). These dimensions can be bounded analytically, according to a method described in Appendix A.

In the case that the $\{\Theta_k\}$ do not commute, the critical submanifold $\mathcal{M}_k^i$ of observable $\Theta_k$ does not necessarily intersect submanifold $\mathcal{M}_{k'}^j$ of an observable $\Theta_{k'}$ that is diagonal in a different basis even if $\mathcal{D}_k^i \cap \mathcal{D}_{k'}^j \neq \emptyset$. In this case, a simple analytical criterion analogous to Lemma 1 for assessing whether (\ref{Eflow}) converges to a Pareto optimum does not exist, but a method presented in Appendix A may be used to determine the maximum possible dimension of the intersection sets.

The most common technique for sampling the Pareto front of multiobjective control problems is to run many independent
maximizations of a cost functional like (\ref{multi}) on the domain $\varepsilon(t)$, using different coefficients
$\{\alpha_k\}$.  The results of Lemma 1 and Theorem 1 indicate conditions under which this strategy may be successful in quantum Pareto optimal control, and demonstrate that it may not always constitute a viable method.  Even when the global optimum of a multiobjective cost function intersects the Pareto front, it is not necessarily advisable to employ such a function for Pareto front sampling on $\varepsilon(t)$, due to the considerable expense of each optimization on this high-dimensional domain \cite{RajWu2008}.

An alternative, generic approach to accelerating Pareto front sampling is to employ stochastic search algorithms that
do not rely on optimization of a scalar objective function.  In recent years, several such multiobjective genetic or
evolutionary algorithms (MOEAs)  - including the elitist nondominated sorted genetic algorithm (NSGA-II)
\cite{Deb2002}, the Pareto-archived evolution strategy (PAES) \cite{Knowles1999} and strength-Pareto evolutionary
algorithm (SPEA) \cite{Zitzer1998} -  have been designed for the problem of multiobjective Pareto front sampling. These algorithms successively sort populations based on nondominance. The NSGA-II algorithm was applied successfully to the
two-objective problem of optimal quantum dynamical discrimination \cite{Wolf2007}. While promising, MOEAs share several drawbacks, most notably a computational complexity that scales as $\mathcal{O}(md^2)$ or $\mathcal{O}(md^3)$, where $d$ denotes the population size. For problems involving control of a large number of observables, these algorithms - in
addition to being less precise than deterministic algorithms - become very expensive since large population sizes are
required to adequately sample the front.

Due to the fact that the quantum control cost functionals (\ref{obs}) are explicit functions of the final unitary
propagator $U(T)$, but only implicit functions of the control field $\varepsilon(t)$, there exists a convenient
alternative to either of the above Pareto optimization strategies based on a hybrid approach that combines kinematic
and dynamic sampling. Assuming full controllability \cite{Ramakrishna1995}, there is at least one  $\e(t)$ in
$\mathcal{P}_w^{\e}$ that maps to any given $U(T)$ in $\mathcal{P}_w^U$. The distribution of $\mathcal{P}_w^U$ in the
domain $\U(N)$ is determined solely by the properties of $\rho(0)$ and the $\{\Theta_k\}$. Therefore, provided that the initial density matrix $\rho(0)$ is known, it is possible to sample Pareto optima on $\U(N)$; this sampling on a
finite-dimensional space will be much more efficient than direct sampling of $\mathcal{P}_w^{\e}$.  Once target points
in $\mathcal{P}_w^U$ have been numerically identified, one may experimentally track a direct path to the corresponding
points in $\mathcal{P}_w^{\e}$, using techniques that will be described in  Section \ref{orthog}.

For example, many different coefficient sets $\{\alpha_k\}$ may be rapidly sampled using the kinematic gradient flow
\cite{RajWu2008} of the multiobservable objective function (\ref{multi}), namely
\begin{equation}\label{kinmult}
\frac{\dd U}{\dd s} = \sum_{k=1}^m
\alpha_k[\Theta_k,U\rho(0)U^{\dag}]U,
\end{equation}
instead of the dynamic flow (\ref{Eflow}). In the case that this flow converges to a (weak) Pareto optimum, it is a very convenient means of sampling segments of the front due to its speed. Of course, single observable kinematic gradient flows may always be used to locate a weakly Pareto optimal point, but do not provide the freedom to preferentially weight the importance of the other observables. Dividing the $m$ observables into mutually commuting sets and applying the weighted flow (\ref{kinmult}) for each set, with the $\{\alpha_k\}$ determined by solving the system of inequalities (\ref{inequalities}), provides a simple and rapid means of sampling the (weak) Pareto optima associated with a specified combination (\ref{condset}) of multiple expectation value constraints.

In the more general case where the $\{\Theta_k\}$ cannot be conveniently subdivided into mutually commuting subsets,
the analytical results above on the distribution of Pareto optima are not immediately useful, but it is straightforward to determine numerically whether certain types of weak and strong Pareto optima exist.
Given an estimate for $\rho(0)$, and the expectation values of $m$ observables at time $T$
specified as the control targets, the submanifold of unitary
propagators $U(T)$ within which the actual propagator may lie is
defined by the system of equations
\begin{eqnarray}\label{pemsystem}
\tr(U(T)\rho(0) U^{\dag}(T) \Theta_k) &=& \chi_k,\quad k=1,...,m,\\  
\left[U^{\dag}(T)U(T)\right]_{ij} &=& \delta_{ij},\quad i,j=1,...,N,
\end{eqnarray}
where $\chi_k$ is the desired expectation value of $\Theta_k$.  If
$\rho(0)$ is full-rank and nondegenerate, the rank of this system of
equations in the $2N^2$ real variables $\mathrm{Re}\left(U_{ij}\right), \mathrm{Im}\left(U_{ij}\right)$ is maximal, resulting in the smallest possible dimension
of the solution submanifold in $\U(N)$. Note that this
system of equations is overdetermined if
\begin{equation}\label{max m and rank rho}m>2Nn-n^2,\quad n={\rm rank}~\rho(0),
\end{equation}
provided $\{\Theta_k\}$ is linearly independent.
For example, if $\rho(0)$ is a pure state, parameterized by $N$
complex coefficients subject to a normalization constraint, then no
more than $2N-1$ independent observables can be simultaneously driven
to arbitrary expectation values. Multiobservable controllability
amounts to controllability over a subset of the independent
parameters of $\rho(0)$.

It is convenient to apply the principle of entropy maximization (PEM) in choosing a target $W$ from the solution
submanifold, since this identifies a propagator that is most likely,
from the perspective of statistical uncertainty, to be the real
propagator of the system given the measurements made during
multiobservable control. In this approach \cite{Buzek1999}, the von
Neumann entropy
\begin{equation}\label{von Neumann entropy}
S(\rho) = \tr(\rho\log \rho), ~~~~\rho(U(T)) = U(T)\rho(0)U^{\dag}(T),
\end{equation}
is maximized on $\U(N)$ subject to the constraints (\ref{pemsystem}).

The following protocol then constitutes a general strategy for kinematic
sampling of Pareto optima: 1) determine the
maximum and minimum expectation values of each of the individual
observables (by simple matching of the eigenvalues of $\rho(0)$,
$\Theta_k$); 2) choose putative sets of observable expectation
values $\{\chi_k\}$ within these respective ranges (setting $l < m$, $l=m$ of these
expectation values to $\chi_k^{\max}$ for weak Pareto optima, strong Pareto optima
respectively);  3) maximize the von Neumann entropy (using, for example, the simplex or
Newton-Raphson algorithms with the constraints imposed as Lagrange
multipliers), assuming the system of equations (\ref{pemsystem})
above is consistent.

In the next section, we describe experimentally implementable algorithms for
the optimization of multiple observable expectation values on the domain of controls $\e(t)$, which can be
used in conjunction with the above kinematic algorithms to efficiently sample the dynamic Pareto front.

\section{Pareto optimal tracking control}\label{orthog}

Once the target observable expectation values $\chi_k,~k=1,\cdots,m$ on the Pareto front
have been determined, it is possible to identify corresponding Pareto optimal control fields $\varepsilon^*(t)$ through
minimization of an objective function of the following form:
\begin{equation}\label{multi2}
\Phi_{M}'(U) = \sum_{k=1}^m \alpha_k|\Phi_{k}(U)-\chi_k|^2, \quad
\alpha_k > 0
\end{equation}
over $\e(t)$, instead of maximization of (\ref{multi}). However, as described in \cite{RajWu2008}, gradient flow algorithms of this type can be inefficient, especially when the rank of the initial density matrix $\rho(0)$ is high, as it may be in large molecular systems or at elevated temperatures.  Moreover, it is inconvenient to employ optimization of (\ref{multi2}) to systematically investigate families of fields $\e^*(t)$ that differ in other experimentally relevant dynamic properties. In this section, we review and extend the more versatile methodology of multiobservable tracking control (MOTC), which drives the expectation values of $m$ observable operators along predetermined paths to the target $\{\chi_k\}$ \cite{RajWu2008}. Once an $\e^*(t)$ has been found, MOTC can be used to continuously trace the dynamic
front to identify solutions that minimize the expenditure of control resources.

\begin{figure*}\label{schematic}
\centerline{
\includegraphics[width=3.3in,height=2.6in]{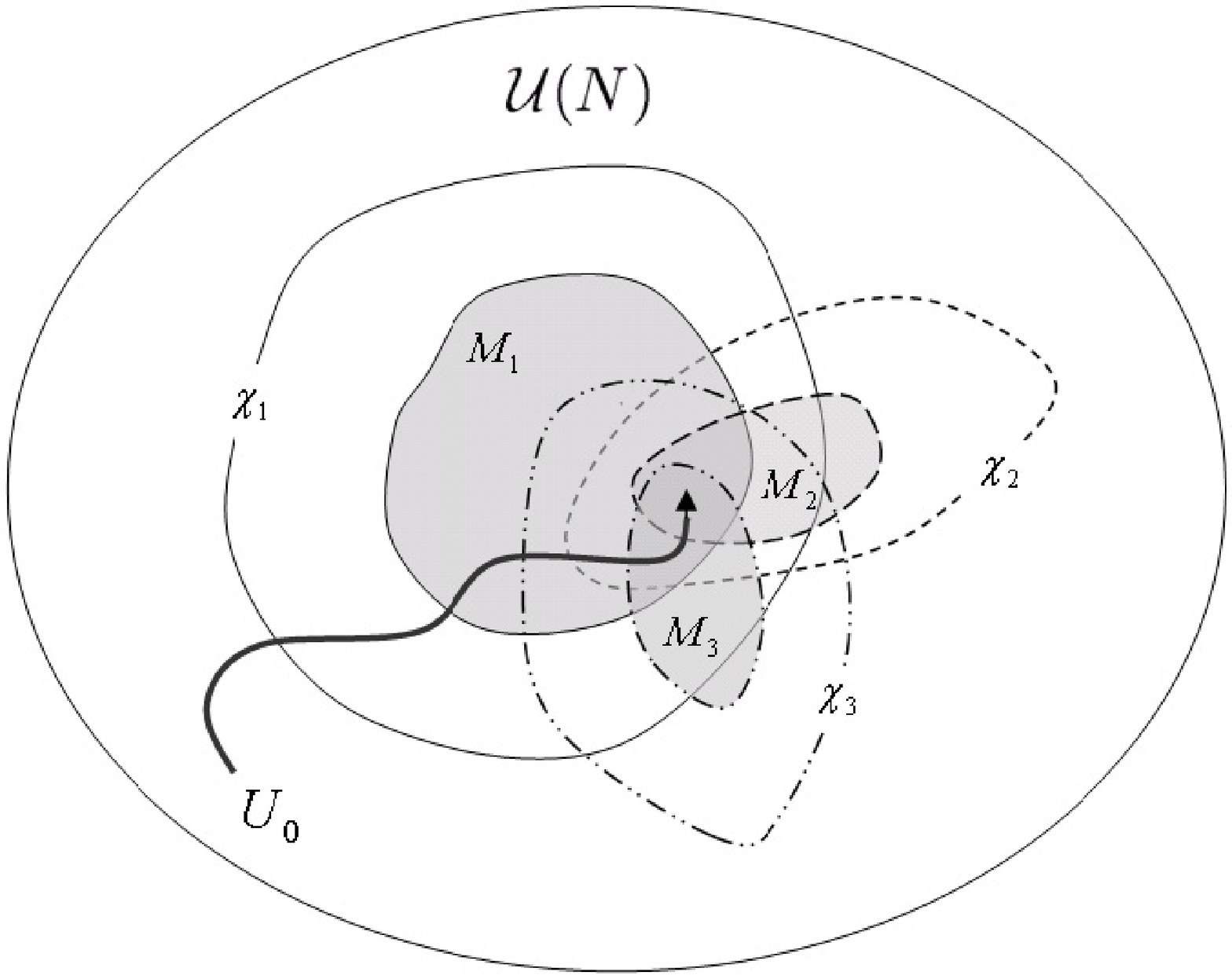}
\includegraphics[width=3.3in,height=2.6in]{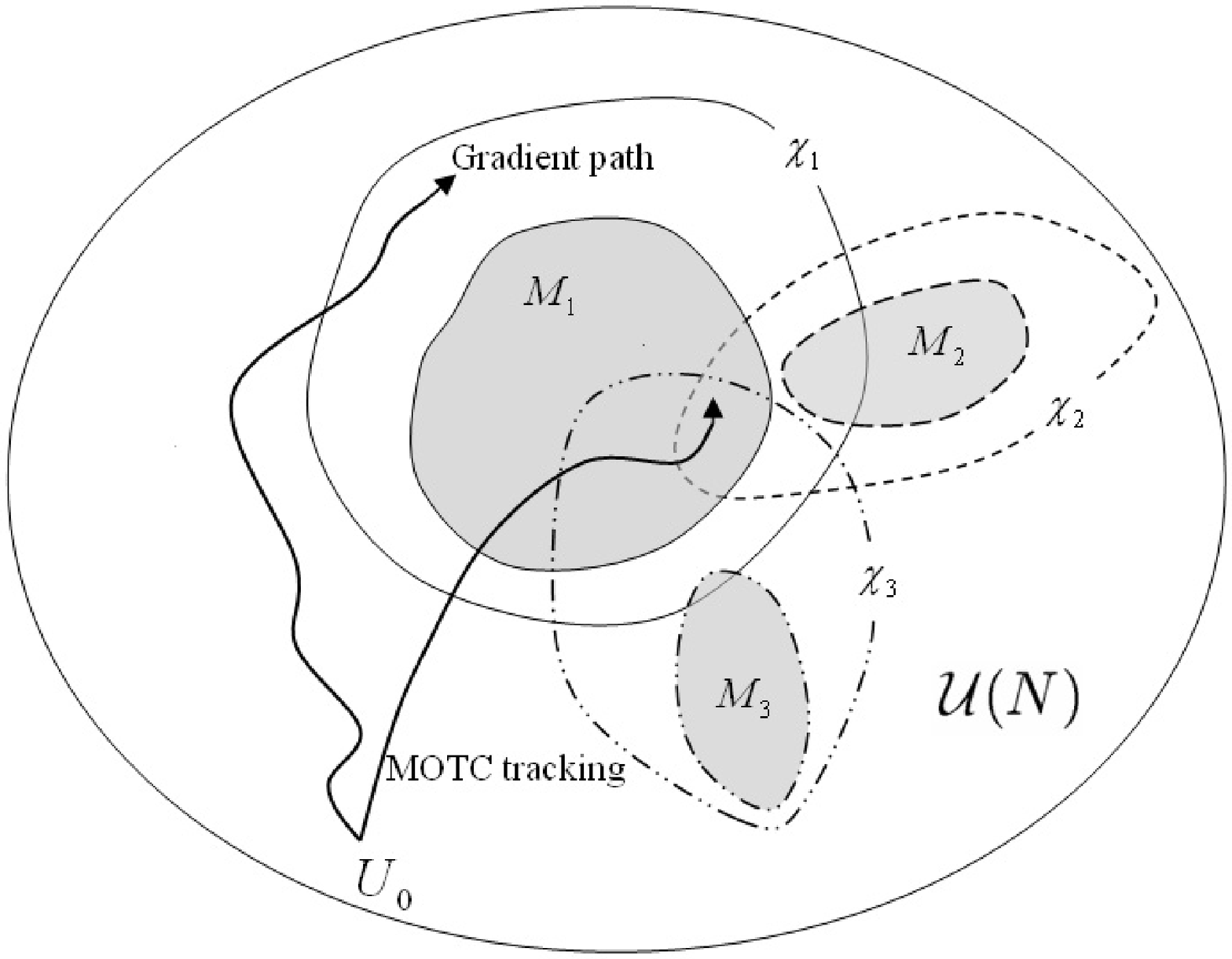}
} \caption{ Quantum Pareto optimal control aims to maximize the
expectation values $\langle \Theta_k\rangle$, $k=1,\cdots,m~$ ($m=3$ in the present case).
For each observable, the expectation value is maximized on a continuous
submanifold $\mathcal{M}_k^{(\max)}$ of unitary propagators $U(T)$, the
dimension of which is determined by the rank and eigenvalue
degeneracies of the initial density matrix $\rho(0)$ and $\Theta_k$.
The observables can be simultaneously maximized if the intersection
$\bigcap_k \mathcal{M}_k^{(\max)}$ is nonempty and a point in the intersection
can be reached under some time-dependent field $\e(t)$ (left panel). This intersection is
a subset of the set of strongly Pareto optimal propagators $\mathcal{P}^U_s$, also called the Pareto frontier.
Depending on $\rho(0)$ and $\{\Theta_k\}$, some problems may not have a completely optimal solution, and may involve a
tradeoff between maximization of different observables. The set $\mathcal{P}^U_w$ of weakly Pareto optimal propagators
consists of the union of intersections of the various $\mathcal{M}_k^{(\max)}$ with level sets
$\{U(T)~|~\langle\Theta_k(T)\rangle =\chi_k\}$
of other observables (right panel). Optimization of multiobjective scalar cost functionals on the domain $\e(\cdot)$
(for example through gradient flows) does not generally lead to Pareto optima. Pareto optimal tracking control involves
computational identification of points in $\mathcal{P}^U_s$ or $\mathcal{P}^U_w$ and subsequent experimental tracking
of a direct path to that manifold on $\e(\cdot)$.}
\end{figure*}

We assume, without loss of generality, that the observables $\Theta_1,...,\Theta_m$
measured at each step of the search are linearly independent. Let $\Phi_s^1,\cdots, \Phi_s^m$ denote the
expectation value paths (functions of an algorithmic parameter $s$) for each observable, which together constitute a
desired path $\textbf{w}_s$ of $\vec \Phi$ in the vector space of multiobservable expectation values.
Then, the change in the control field along the track $\frac{\partial \e_s(t)}{\partial s}$
must satisfy the following condition:
\begin{equation}\label{gendiff2}
\frac{\dd \Phi_s^i}{\dd s} = \int_0^T \frac{\delta \Phi_s^i}{\delta
\e_s(t)}\frac{\partial{\e_s(t)}}{\partial s}\dd t=\frac{\dd \textbf{w}_s^i}{\dd s}.
\end{equation}
where $\Phi_i=
\tr(U(T)\rho(0)U^\dag(T)\Theta_i)$
This is a Fredholm integral eqn of the first kind for
the unknown partial derivative $\frac{\partial{\e_s(t)}}{\partial s}$, given $\e_s(t)$ at $s$ and all $t \in [0,T]$.
To solve for the algorithmic flow $\frac{\partial \e_s(t)}{\partial s}$ that
tracks $\textbf{w}_s$, it is necessary to expand it on a basis of independent functions.
It is convenient to make this expansion on the basis of the independent observable
expectation value gradients:
\begin{equation}\label{expansion}
\frac{\partial \e_s(t)}{\partial
s} = \sum_{j=1}^m x^{j}_s \frac{\delta \Phi_s^j}{\delta
\e_s(t)}+f_s(t),
\end{equation}
where $f_s(t)$ contains the degrees of freedom outside the subspace spanned by
$\frac{\delta \Phi_s^j}{\delta \varepsilon_s(t)}$. As described in \cite{RajWu2008}, $f_s(t)$ can take on a variety of
forms depending on auxiliary dynamical costs on $\e(t)$ that are to be minimized.
Inserting the expansion (\ref{expansion}) into the integral equation (\ref{gendiff2}), we have
$$\sum_{j=1}^m x^{j}_s \int_0^T\frac{\delta \Phi_s^i}{\delta \e_s(t)}~
\frac{\delta \Phi_s^j}{\delta \e_s(t)}\dd t= \frac{\dd
\textbf{w}^{i}_s}{\dd s}-\int_0^T \frac{\delta \Phi_s^i}{\delta
\e_s(t)}f_s(t)\dd t.$$
Defining the $m$-dimensional vector
$\textbf{a}_s(t)=(a_s^1(t),\cdots,a_s^m(t))$ by
\begin{equation*}
a_s^i(t)=\frac{\delta \Phi_s^i}{\delta \e_s(t)}=-\frac{\imath}{\hbar}\tr\left(\rho(0)
\big[U^{\dag}_s(T)\Theta_iU_s(T),\mu_s(t)\big]\right),
\end{equation*}
and the MOTC Gramian matrix $\Gamma_s$ by
\begin{equation}\label{Gramian}
(\Gamma_s)_{ij} = \int_0^T
a^i_s(t')a^j_s(t')\dd t',
\end{equation}
it can then be shown \cite{RajWu2008} that the initial value problem for $\e_s(t)$ given
$\e_0(t)$ is:
\begin{equation}\label{vectrack}
\frac {\partial \e_s(t)}{\partial s} = f_s(t) + \left[\frac{\dd
\textbf{w}}{\dd s}-\int_0^T\textbf{a}_s(t')f_s(t')\dd t'
\right]^T\Gamma_s^{-1}\textbf{a}_s(t).
\end{equation}
This expression can be integrated either numerically (using standard methods described in
\cite{RajWu2008}) or experimentally (Section V) to track the desired path $\textbf{w}_s$ to the dynamic Pareto front.
The invertibility of the Gramian matrix $\Gamma_s$ corresponds to the
ability to move from the point $\vec \Phi_s$ in multiobservable space to any infinitesimally close point $\vec \Phi_{s+\dd s}$ through an infinitesimal change $\delta \e_s(t)$ in the control field \cite{RajWu2008}.
With minor modifications \cite{RajWu2008}, the MOTC tracking differential equation (\ref{vectrack}) can
also be used to carry out multiobjective optimizations for non-observable objectives.

The MOTC algorithm can be applied to follow any arbitrary vector track $\textbf{w}_s$ of
observable expectation values. It is convenient to choose $\textbf{w}_s$ based on
the corresponding set of possible paths $U_s(T)$ in the domain $\U(N)$ of unitary propagators,
since the latter contains the most information about the dynamics of the system at time $T$.
Let $\mathcal{L}_s=\{V_s ~|~ \Phi_k(V_s) = w_s^k,~k=1,...,m\}$ denote the set of propagators that map
to the observable track at algorithmic time $s$.  Denote by $B_s$ the system of
equations (\ref{pemsystem}) with $\{\chi_k\}$ replaced by $\textbf{w}_s$.
Assuming this system is consistent, the dimension of $\mathcal{L}_s$ is equal to $N^2$ less the number of independent
equations in $B_s$. As such, the path $U_s(T)$ followed by the optimization algorithm in $\U(N)$ will, on average, be
more similar to a target unitary track $Q_s$, for a greater number $m$
of orthogonal observables (note $\mathrm{dim}~\mathcal{L}_s$ also depends on the
eigenvalue spectra of $\rho(0)$ and $\{\Theta_k\}$).
As a function of the algorithmic step $s$, the term $\left[\int_0^T\textbf{a}_s(t')f_s(t')\dd t'
\right]^T\Gamma_s^{-1}$ in equation (\ref{vectrack})
will adjust the step direction so that the possible unitary propagators $V_s$ at each step are
constrained within the subspace $\mathcal{L}_s$ (provided $\Gamma$ is invertible; see Section VI).
In Section \ref{expt}, we describe special choices for $\textbf{w}_s$
that are globally optimal from the standpoint of mean distance traveled in $\U(N)$.

An alternative MOTC-based approach to identifying weak dynamic Pareto optima is
to maximize the expectation values of successive observable operators while holding
those previously maximized at constant values. This requires successively applying $m$ MOTC
searches of dimensions $1,\cdots,m$, respectively; these searches may be carried out in an
order that reflects the relative importance of the observable
objectives. Let $(p_1,\cdots,p_m)$, where $p$ is a permutation on $m$ indices, denote this order and
let $\chi^{p_i}$ denote the conditional maximum of $\Phi^{p_i}(\e(t))$ given $\Phi^{p_j}(\e(t)) = \chi^{p_j},~1\leq j < i$. This approach then amounts to successive integrations of (\ref{vectrack}) with
\begin{eqnarray}\label{sequential}
\textbf{w}_s^{(r)} &=& (\chi^{p_1},\cdots,\chi^{p_{r-1}},w_s^r),\\
\textbf{a}_s^{(r)} &=& (a_s^{p_1},\cdots,a_s^{p_r}), \quad r=1,\cdots,m,
\end{eqnarray}
where $w_s^r$ is any monotonically increasing function of $s$.
For maximization of $m$ observables, there are $m!$ such combinatorially distinct
control strategies.

The final observable expectation values $\{\chi_k\}$ will seldom be the only properties of a control solution that are
relevant experimentally. In general, other dynamic properties of the field, such as its total fluence $\int_0^T \e^2(t)
dt$, will have implications for the feasibility of a control experiment. Hence, the ability to continuously traverse a
submanifold of the dynamic Pareto front, once it has been found, may be essential for identifying a Pareto optimal
control $\e^*(t)$ with the most desirable physical properties.
There generally exists an infinite number of control field solutions corresponding to any given
combination of final observable expectation values $\{\chi_k\}$, as evidenced by the presence of the free function
$f_s(t)$ in equation (\ref{vectrack}). The MOTC algorithm offers a systematic means of exploring these families of
solutions through specification of $f_s(t)$.
To maintain the objective function values $\{\Phi_k\}$ at their target values $\{\chi_k\}$ while sampling fields
corresponding to a free function $f_s(t)$, the following initial value problem for $\e_s(t)$ is solved:
\begin{equation}\label{leveltrack}
\frac {\partial \e_s(t)}{\partial s} = f_s(t) - \left[\int_0^T\textbf{a}_s(t')f_s(t')\dd t'
\right]^T\Gamma_s^{-1}\textbf{a}_s(t).
\end{equation}
With the observable expectation values fixed, different choices of the free function $f_s(t)$ correspond to different
trajectories through the associated portion of the dynamic Pareto front. The final unitary propagator $U(T)$ may
change during such excursions. Choosing $f_s(t)=- \frac{1}{\eta} \frac{\e_s(t)}{S(t)}$ , where $S(t)>0$ is an arbitrary weight function (e.g., a Gaussian) and $\eta$ is a constant that controls numerical instabilities, will minimize fluence at each
algorithmic time step $s$ \cite{Rothman2006b}. Note that equation (\ref{leveltrack}) can also be used to identify
fluence-minimizing controls for other multiobservable control problems like optimal state preparation.

The connectedness of the submanifold of $\mathcal{P}_w^{\e}$ corresponding to the expectation values $\{\chi_k\}$
determines which $\e^*(t)$ can be accessed by application of (\ref{leveltrack}). It has been shown \cite{Pechen2008}
that the kinematic level sets $\{U(T)~|~\Phi_k(U(T)) = \chi_k\}$ of single observables are connected manifolds.
However, the dynamic level sets $\{\e(t)~|~\Phi_k(\varepsilon(t)) = \chi_k\}$ may under certain circumstances
\cite{Dominy2008} be disconnected. Moreover, it is possible that the intersections
of the kinematic level sets of multiple observables (i.e., the solutions to the system of equations (\ref{pemsystem}))
consist of disconnected submanifolds. If this intersection is disconnected, it follows that the dynamical
intersection set is also disconnected \cite{Dominy2008}.

A disadvantage of the purely dynamic Pareto front sampling strategy (\ref{sequential}) is that since
the manifold of control fields $\{\e(t)~|~ \Phi_k(\e(t))=\chi_k,~1\leq k \leq m\}$ to which
the algorithm converges may in general consist of disconnected submanifolds,
certain classes of solutions may not be accessible from a canonical
starting field $\e^*(t)$ by the homotopy tracking algorithm (\ref{leveltrack}).
By contrast, sampling of the kinematic Pareto front by the methods described in Section \ref{kinpareto}
can (at the expense of additional initial overhead required for state estimation) identify propagators $U^*(T)$ that
lie in different disconnected submanifolds of the front,
such that (\ref{leveltrack}) can subsequently be applied to reach any control field in the local
vicinity of the canonical field $\varepsilon^*(t)$ obtained from application of (\ref{vectrack}).
On the other hand, a virtue of the method (\ref{sequential}) is that it is also applicable to quantum Pareto optimal control problems other than observable maximization, where kinematic identification of
Pareto optima may not be possible.

Maximization of (\ref{multi2}) or of (\ref{multi}) (assuming $\{\alpha_k\}$ are restricted to the ranges stipulated by
the system of inequalities (\ref{inequalities})), with various coefficient sets $\{\alpha_k\}$, will also lead to
distinct points on the dynamic Pareto front with the same observable expectation values but different $U(T)$ and/or
dynamic properties of the field. However, such a strategy is far less efficient than applying equation
(\ref{leveltrack}), since a separate optimization must be carried out for each set of coefficients. Further
mathematical details on the formal correspondence between these two approaches, from the perspective of differential
geometry, may be found in the treatise \cite{Hillermeier2001}.

The observables  $\{\Theta_k\}$ of interest are not necessarily
orthogonal. However, the statistical efficiency of MOTC (i.e., the
tracking accuracy for a given number of measurements) is easiest to
assess when orthogonal observables are measured at each step. For
simplicity, we represent each of the Hermitian matrices as an
$N^2$-dimensional vector with real coefficients. Then Gram-Schmidt
orthogonalization offers a convenient means of constructing an
orthogonal basis of $m$ linearly independent $N^2$-dimensional
vectors, $\Theta'_1(T),...,\Theta'_m(T)$ that spans the associated
subspace - any element of the set $\{\Theta_k\}$ can be expressed as
a linear combination of the basis operators in this set, i.e., for
any $k$, $\Theta_k = \sum_{i=1}^m c_{ki} \Theta'_i$. When the
desired tracks of the target observable expectation values are thus
expanded on a set of experimentally measured orthogonal observables,
we have
\begin{equation}\label{multgrad}
\frac{\delta \Phi_k}{\delta \e_s(t)} = -\frac{\imath}{\hbar}\sum_i
c_{ki}\tr\left( \left[\Theta_{i}'(T),\rho(0)\right]\mu(t)\right)
\end{equation}
for the gradient of each of the observable expectation values
$\langle \Theta_k \rangle$ with respect to the control.

\section{Experimental implementation}\label{expt}

This section discusses methodologies for the laboratory implementation of MOTC and Pareto optimal control.
Laboratory implementation of quantum optimal control is typically carried out in open loop with numerical
search algorithms guiding the experimental control field updates (e.g., via a laser pulse shaper) at each algorithmic
step \cite{Levis2001}.

\subsection{Initial state estimation}

In order to kinematically sample the Pareto front as described
above, it is necessary to have an estimate for the initial density
matrix $\rho(0)$. In the case that $\rho(0)$ is not a thermal mixed
state, it can be determined by the method of maximal likelihood
estimation (MLE) of quantum states \cite{Banaszek1999}. Here, the
likelihood function
$$L(\hat \rho) = \prod_{i=1}^{n} \tr(\hat \rho(0) \mathcal{F}_i),$$
where $\mathcal{F}_i$ denotes a positive operator
valued measure (i.e., one of a complete set of $N^2-1$ orthonormal observable operators) corresponding to the outcome of the $i$-th measurement, describes the probability of obtaining the set of $n$ observed outcomes for a given trial density matrix $\hat \rho$. This likelihood function must be maximized over the set of admissible density matrices. An effective parameterization of $\hat \rho$ is the Cholesky decomposition $\hat
\rho(0) = \hat T^{\dag} \hat T$ -- where $T$ is a complex lower
triangular matrix with real elements along the diagonal -- which
guarantees positivity and Hermiticity. The remaining condition of unit
trace is imposed via a Lagrange multiplier $\lambda$. Maximization of the function
$$\mathcal{L}(\hat T) = \sum_{i=1}^n \ln \tr (\hat T^{\dag} \hat T \mathcal{F}_i) - \lambda \tr(\hat
T^{\dag}\hat T),$$
with the choice $\lambda=n$ for the Lagrange multiplier, guarantees normalization of $\hat \rho$ at the
MLE estimate \cite{Banaszek1999}.
Standard numerical techniques, such as Newton-Raphson or uphill
simplex algorithms, are used to search for the maximum of $\mathcal{L}$ over the
$N^2$ parameters of the matrix $\hat T$.

\subsection{Minimal-length $\textbf{w}_s$ and error correction}

There are many possible choices for the track $\textbf{w}_s$ that leads to a point $W$ in
$\mathcal{P}_w^U$. However, it is desirable to choose paths
that are globally optimal from a geometric perspective. The
Riemannian Bures metric on the space of density matrices
\cite{Uhlmann1976} could be used for this purpose, but this metric
cannot be expressed in compact form for arbitrary Hilbert space
dimensions, especially for arbitrary $m$-dimensional subspaces of
the Hilbert space. We thus adopt the more convenient approach of
defining the target multiobservable track as the
$\textbf{w}_s$ that maps to the geodesic $Q_s$ in $\U(N)$ between
any $U_0$ consistent with the initial control guess and $W$, i.e.,
\begin{equation}\label{geotrack}
\textbf{w}_s^k \equiv \sum_ic_{ki}\tr\left\{\rho(0) Q_s^{\dag}
\Theta_i'
 Q_s\right\},
\end{equation}
where $Q_s=U_0\exp(\imath As)$ with $A = -\imath\log(W^{\dag}U_0)$.

Error-correction methods can be applied to
account for deviations from the track $\textbf{w}_s$. These generally
involve the addition of a correction term $\mathbf{c}_s$
to the tracking differential equation (\ref{vectrack}) such that
\begin{widetext}
\begin{equation}\label{motcerr}
\frac {\partial \e_s(t)}{\partial s}= f_s(t) +
\left[\mathbf{c}_s+\frac{\dd \textbf{w}_s}{\dd s}-
\int_0^T\textbf{a}_s(t')f_s(t')\dd
t'\right]^T\Gamma_s^{-1}\textbf{a}_s(t),
\end{equation}
\end{widetext}
where $\mathbf{c}_s$ may be taken to be a scalar multiple of the difference
between the actual vector of observable expectation values and the target track, i.e.,
$$\mathbf{c}_s=\beta(\mathbf{w}_s-\vec\Phi_s),~~~\beta>0.$$
Note that since this error-correction method requires only estimation of the
expectation values of the $m$ observable operators $\{\Theta_k\}$, it is
straightforward to implement in an experimental setting.

Runge-Kutta (RK) integration of the MOTC differential equation (\ref{motcerr}) can further decrease tracking errors by
employing derivative information at multiple points across the algorithmic step, instead of only at a single point. In
principle, RK may be applied experimentally at a substantially lower cost than including second order functional
derivatives in the tracking differential equations, since the latter requires estimation of the Hessian matrix of the
observable expectation values with respect to the control. Moreover, although it is possible to implement second-order
MOTC numerically, accurate experimental estimation of the Hessian is difficult due to the inevitable presence of
experimental noise. Thus, RK integration is the preferred method for further improvement of the experimental accuracy
of MOTC algorithms.

\subsection{Choice of measurement bases}

Gram-Schmidt orthogonalization, as discussed in the previous
section, is a convenient method for obtaining a canonical orthogonal
observable basis for MOTC, but from the point of view of experimental overhead,
it is generally not statistically efficient to measure these
observables. The complete set of $N^2-1$ possible orthogonal
observables can instead be expanded on $N+1$ orthonormal measurement
bases. Here, we use mutually unbiased measurement bases (MUB)
\cite{Wootters1981}, which are known to provide tight confidence
intervals on the expectation values of a set of multiple observables
for a finite number of measurements (the advantage in statistical
efficiency increases with $m$). The expression for MUB differs based
on the Hilbert space dimension $N$. When $N$ is the power of an odd
prime (which encompasses all cases considered in the current work),
these bases $V^{(r)}$ are given by

\begin{equation*}\label{mubbases}
V_{pq}^{(r)}=\left\{
\begin{array}{ll}
\delta_{pq},& r=0\\
\\
\frac{1}{\sqrt N}\exp[\frac{2\pi i}{N}(rq^2+pq)],~~&1\leq r\leq N.\\
\end{array}\right.
\end{equation*}
The orthonormal observables $\Theta'$ in equation (\ref{multgrad}) can then be taken to be
\begin{eqnarray}\label{mub}
\Theta_{r(N-1)+i}' &=& V^{(r)}\tilde \Theta_i'(V^{(r)})^{\dag},\\
\tilde\Theta_i' &=& |i\rangle\langle i| = diag\{0,...,1,...0\},
\end{eqnarray}
for $1 \leq i \leq N-1$, $0 \leq r \leq N$. However, the orthonormal observables in these bases need not be projection operators.

Every quantum measurement in a basis $V^{(r)}$ provides information
regarding the parameters $p_1,...,p_{N}$ of the multinomial
distribution corresponding to that basis ($N-1$ of these may be considered parameters of $\rho(T)$). The measurement of an observable $\Theta$ with $s$ distinct eigenvalues returns $s-1$
\textit{independent} parameters from the set. Control of multiple observables
can be achieved for the lowest experimental cost by measuring
nondegenerate observables in the bases of interest.  Moreover,
tight bounds on those gradient estimates will be obtained if the
measurement bases are mutually unbiased.
Therefore, for each MUB basis, assume a full-rank, nondegenerate observable $\Theta$ is measured that is diagonal
in that basis, i.e., $\Theta = V^{(r)} \Sigma (V^{(r)})^{\dag}, \quad \Sigma = diag\{\sigma_1,\cdots,\sigma_N\}$.
The multinomial distribution parameters $p_1,...,p_{N-1}$ are then formally given by
\begin{equation*}
p_i = \tr\left(\rho(T)V^{(r)}\tilde\Theta_i'(V^{(r)})^{\dag}\right),~1 \leq i \leq N-1,
\end{equation*}
with the remaining $p_N = 1 - \sum_{i=1}^{N-1} p_i$. These correspond experimentally to the frequencies with which the measurement outcomes $\sigma_i$ are observed.
Assuming for simplicity that the measurement basis $V^{(r)}$ has
been chosen such that it coincides with a basis within which one of
the $m$ target observables $\Theta_k$ is diagonal (extension to the more general case is
straightforward), and writing
$\Theta_k = V^{(r)} \Xi (V^{(r)})^{\dag}$ where $\Xi =
diag\{\gamma_1,\cdots,\gamma_N\}$, the expectation values of the $m$
target observables at each MOTC step can then be written in terms of the experimentally estimated frequencies as
$$\langle\Theta_k\rangle = \sum_{i=1}^N c_{ki} p_i \gamma_i,$$
where the $c_{ki}$ are the expansion coefficients in (\ref{multgrad}). This approach is generally more efficient than direct estimation of the $\langle \Theta_k\rangle$.

\subsection{Gradient estimation}

In an experimental setting, the gradients $\frac{\delta \Phi_k}{\delta \e_s(t)}$
may be estimated by statistical sampling of points near the current
control field $\e_s(t)$ \cite{Roslund2008} instead of the method of finite differences;
the latter is inaccurate due to the inevitable presence of
noise (e.g., laser noise) in the laboratory. Let us denote a discretized representation
of the control field $\e(t)$ by the vector $\vec{x}$ (e.g., the 128 pixel
settings in a laser pulse shaper) with associated domain $\mathcal{D}$, the
point at which the gradient is evaluated by $\vec{x}_0$, and a
symmetric distribution function around $\vec x_0$ over
$\mathcal{D}$ by $\pi$.

It can be shown \cite{Roslund2008} that the following expression constitutes a second-order approximation to $\nabla \Phi(\vec{x}_0)$:
$$\nabla\Phi(\vec x_0)\approx {\Sigma}^{-1}\int_{\mathcal{D}}(\vec x-\vec x_0)\Phi(\vec x)\pi(\vec x-\vec
x_0)\dd \vec x,$$where the covariance matrix is
$${\Sigma}=\int_{\mathcal{D}}(\vec x-\vec x_0)(\vec x-\vec x_0)^T\pi(\vec x-\vec x_0)\dd \vec x.$$
This expression can be evaluated via Monte Carlo integration over $N_{\exp}$ measurement results:
$$\nabla \Phi(\vec{x}_0) \approx \frac{{\Sigma}^{-1}}{N_{\exp}}\sum_{i=1}^{N_{\exp}}(\vec x^{(i)}-\vec x_0)\Phi(\vec x^{(i)})\pi(\vec x^{(i)}-\vec
x_0).$$
Open-loop steepest ascent control field search using a femtosecond laser pulse shaper, with the gradient evaluated at each iteration via this technique, successfully
reached the global maximum of the control landscape for second harmonic generation \cite{Roslund2008}, indicating a
robustness of such gradient-based search algorithms to noise.

\begin{figure}
\centerline{
\includegraphics[width=5.7in,height=4.1in]{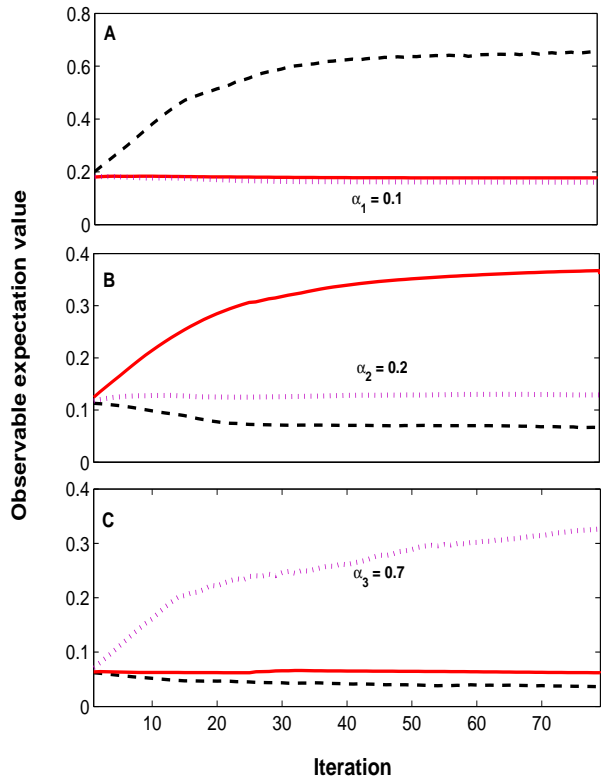}
}
\caption{MOTC-based Pareto optimum ($\mathcal{P}_w^{\e}$) sampling for three observables in an 11-level system. Weakly
Pareto optimal unitary matrices were determined using the kinematic gradient (\ref{kinmult}) followed by MOTC to
identify optimal controls. The convex weights in equation (\ref{multi})
were set to $\alpha_1= 0.7,~\alpha_2= 0.2, \alpha_3= 0.1$ (dashed);
$\alpha_1= 0.2,~\alpha_2= 0.7, \alpha_3= 0.1$ (solid); $\alpha_1= 0.1,~\alpha_2= 0.2, \alpha_3= 0.7$ (dotted).
$\rho(0)$ was a thermal mixed state. \textbf{A)} Observable $\Theta_1$: rank three diagonal degenerate observable with the three
highest energy levels targeted; \textbf{B)} Observable $\Theta_2 = |2\rangle\langle 2|$; \textbf{C)} Observable
$\Theta_3 = |3\rangle\langle 3|$.
}\label{pareto}
\end{figure}

\section{Examples}\label{numerical}

As a representative Pareto optimal control scenario, we apply tracking control to the weighted maximization of three mutually commuting observables. In addition, in order to illustrate the characteristic features of problems - such as optimal mixed state preparation - that require the control of $m > 2N-1$ observables, we apply MOTC using multiple mutually unbiased measurement bases to a multiobservable control problem involving control of 40 observables in an 11-level system.

The tracking examples below employ an 11-dimensional Hamiltonian of the form (\ref{ham}), with 
\begin{eqnarray}\label{system example}
H_0 &=& diag~\{0.1,0.2,\cdots,1.1\},\\
\mu_{ij}&=&\left\{
\begin{array}{cl}
1,& i=j;\\
0.15,&|i-j|=1;\\
0.08,&|i-j|=2;\\
0, &\text{otherwise}.
\end{array}\right.
\end{eqnarray} It can verified (by checking that the rank of the Lie algebra generated by and $H_0$, $\mu$ equals $N^2$), that this system is fully (propagator) controllable \cite{Ramakrishna1995, Albertini2003}.
MOTC and multiobservable gradient flow algorithms were implemented using the numerical methods described in
\cite{RajWu2008}.

\subsection{Quantum control landscape Pareto front sampling}

\begin{figure}
\centerline{
\includegraphics[width=6.5in,height=5.7in]{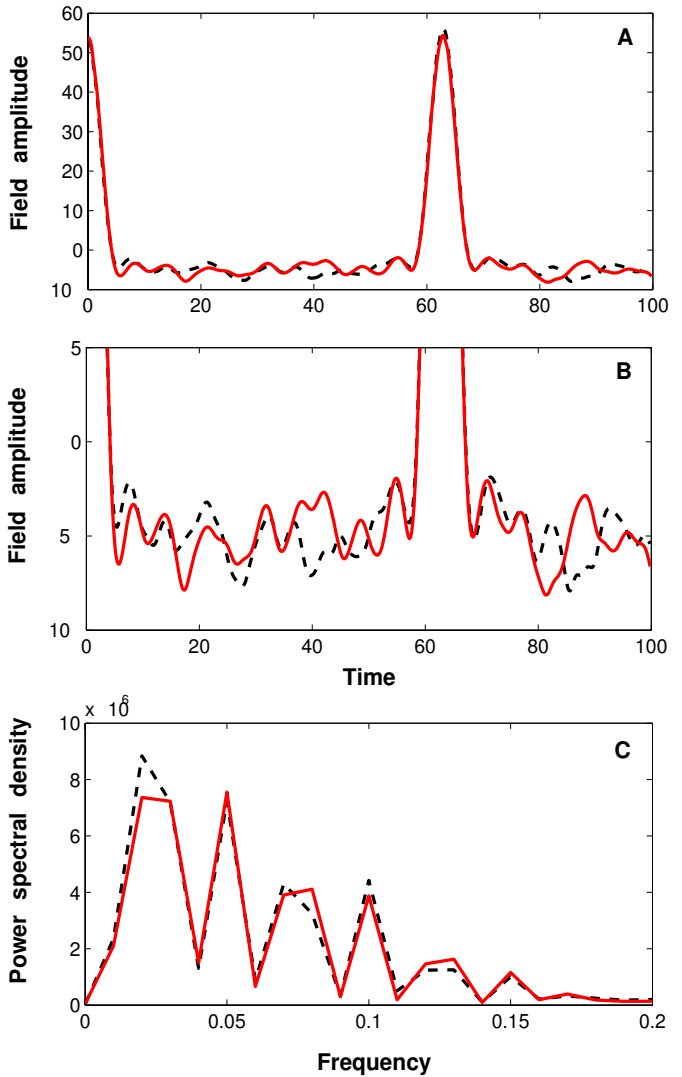}
} \caption{Optimal control fields \textbf{A)}, \textbf{B)} (magnified) and their Fourier power
spectra \textbf{C)} obtained by the MOTC Pareto optimum sampling depicted in
Fig. \ref{pareto}, for weights $\alpha_1=0.7, ~\alpha_2=0.2,
~\alpha_3=0.1$ (dashed) and
$\alpha_1=0.2,~\alpha_2=0.7,~\alpha_3=0.1$ (solid).
$\varepsilon_0(t)$ was of the form (\ref{randfield}) with amplitudes
weighted according to the number of times the frequency $\omega_{ij}$
appears in the transition frequency spectrum.
}\label{paretofield}
\end{figure}

A common objective in quantum multiobservable maximization problems is the identification of control fields that
maximize a single observable expectation value while selectively weighting the importance of increasing the expectation values of several auxiliary observables. As discussed above, such control fields are weakly Pareto optimal. These weak
Pareto optima can be efficiently sampled using the methods described in Sections III and IV.
As an example of weighted multiobservable maximization, several combinations of convex weights in multiobservable
objective function (\ref{multi}) were employed in separate trials, representing desired trends in the relative
magnitudes of the respective observable expectation values of three commuting observables;
each observable was assigned the highest weight in several trials. The weakly Pareto optimal set corresponds in this
case to points on $\varepsilon(t)$ where at least one of the 3 observable expectation values is at its maximum possible value; for the purposes of numerical illustration,  $\sim 85\%$ of the maximum achievable value was considered
sufficient. The kinematic gradient (\ref{kinmult}) was applied to maximize the multiobservable objective function on
the domain $\U(N)$; in the case of these observables, it always reached a weakly Pareto optimal point. The expectation
value of the observable assigned the highest weight reached the greatest value.

MOTC (equation (\ref{vectrack})) was then applied to track multiobservable expectation value
paths (\ref{geotrack}) to controls in $\mathcal{P}_w^{\e}$ that map to these points in $\mathcal{P}_w^U$ (Fig.
\ref{pareto}). Fig. \ref{paretofield}A compares the optimal control fields obtained by MOTC and their Fourier power
spectra at a  point on the weak Pareto optimum where observable 1 dominates, with those at
a corresponding point where observable 2 dominates.
These points do not lie on the same level set of any observable. Nonetheless, note that their Fourier power spectra
(Fig. 3B) are very similar, indicating that many similar fields reside at the distinct weakly Pareto
optimal manifolds corresponding to different dominant observables.
By contrast, optimal fields that lie on the level set of a single observable may display very different Fourier
spectra, if different numbers $m$ of observables are controlled (as shown in Fig. \ref{fieldevol} below).
This suggests that the major contributor to the complexity of optimal fields in Pareto optimal
control problems is the number of parameters of the density matrix or $U(T)$ that are simultaneously constrained, and
not the tradeoff between the expectation values of the various observables on the Pareto front.

In order to assess the advantage of precomputing a propagator in
$\mathcal{P}_w^{U}$ and tracking to a corresponding point in $\mathcal{P}_w^{\e}$ via MOTC, the
multiobservable dynamical gradient (\ref{gradmult}) was applied directly in an adaptive step
size (line search) steepest ascent algorithm. Although the multiobservable gradient flow
converged to a weak Pareto optimum, the dynamical gradient flow was substantially
less efficient in most weighted multiobservable maximizations compared to single observable
maximizations (data not shown). Typically, none of the $m$ observable expectation values increase
monotonically in dynamical multiobservable steepest ascent, whereas
the expectation value of the most highly weighted observable does
rise monotonically in the kinematic multiobservable gradient flow, presumably
due to the lower dimensionality of the search domain. In an experimental setting where
noise may obscure minute differences in observable expectation values, application of
steepest ascent may therefore not be as efficient as it has been found to be
for the control of single observables \cite{Roslund2008}. By contrast, kinematic sampling on $\U(N)$
(after estimation of the state $\rho(0)$) is a reliable and rapid
method of sampling $\mathcal{P}_w^U$.

In the above example, the observables were mutually commuting, i.e.
belonged to the same measurement basis. In this case, the kinematic gradient flow (\ref{kinmult})
could be used to locate weakly Pareto dominant points, since the maximum submanifold
$\mathcal{M}_M^{\max}$ of $\Phi_M$ intersected the maximum $\mathcal{M}_k^{\max}$ of one of the
single observable cost functions $\Phi_k$ (here, the cost functions satisfied the conditions in
Lemma 1 guaranteeing Pareto convergence). In the more general case where the targets are $m$-tuples
of arbitrary observable expectation values, or points on the Pareto front where specific expectation values of
the $m-1$ dominated observables are desired, the von Neumann entropy
(\ref{von Neumann entropy}) can be maximized, instead of employing the kinematic gradient flow.

\subsection{Effect of state preparation}

\begin{figure*}
\centerline{
\includegraphics[width=7.5in,height=3in]{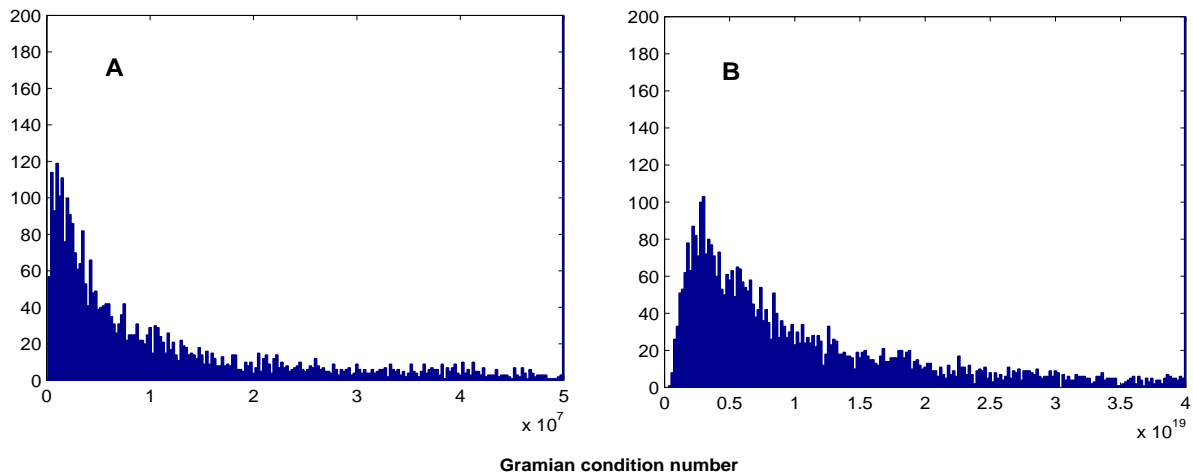}
} \caption{Gramian matrix condition number distributions for an
11-level system (random internal Hamiltonian $H_0$ and dipole
operator $\mu$): multiple measurement bases and the effect of state
preparation. The amplitudes and phases of the modes of the control
field $\varepsilon(t)$ were sampled randomly from the uniform
distributions $(0,1]$ and $(0,2\pi]$, respectively, with the mode frequencies tuned to the
transition frequencies of the system. \textbf{A)} MOTC, 20
observables: pure state $\rho(0)$. \textbf{B)} MOTC, 40 observables:
pure state $\rho(0)$. Observables were drawn sequentially from
successive mutually unbiased measurement bases in equation
(\ref{mub}).
}\label{gramfig2}
\end{figure*}

The MOTC Gramian matrix $\Gamma$ must be invertible at a given control field $\e_s(t)$ in order
for the tracking algorithm to be able to follow any arbitrarily designated path to the Pareto front
$\mathcal{P}_w^{\e}$. Typically, a Gramian condition number $C >
10^9$ results in large numerical errors upon inversion, and would be
expected to compromise the accuracy of tracking, due to the
sparseness of control field increments $\delta \varepsilon(t)$
that are capable of driving the system to the arbitrary neighboring states.

As the number of controlled observables $m$ increases,  the condition number $C$
can rise steeply beyond a critical $m$, where additional observables
cannot be assigned arbitrary local tracks.
Analytically, when the condition (\ref{max m and rank rho}) is
satisfied, the $m$ functions of time $\frac{\delta \Phi_k}{\delta
\varepsilon(t)}$ that dictate local multiobservable controllability
cannot remain linearly independent, resulting in an ill-conditioned
Gramian $\Gamma$. This feature of multiobservable local
controllability is independent of the
Hamiltonian. However, in practice, condition numbers typically do
not explode at a critical $m$, due to numerical inaccuracies in the
singular value decomposition of matrices that are close to singular.

Fig. \ref{gramfig2} compares the $\Gamma$ condition number distributions
for a system with pure $\rho(0)$, for $m_1=20$ and $m_2=40$,
using randomly sampled fields $\varepsilon(t)$ of the form
\begin{equation}\label{randfield}
\varepsilon(t) = \sum_{i=1}^N \sum_{j=i+1}^N A_{ij}\sin\left(\omega_{ij}t+\phi_{ij}\right), \quad 0\leq t \leq T,
\end{equation}
where $\omega_{ij}=|E_i-E_j|$ denote the transition frequencies
between energy levels $E_i,~E_j$ of $H_0$, $\phi_{ij}$ denotes a
phase sampled uniformly within the range $(0,2\pi]$, and $A_{ij}$
denotes a mode amplitude sampled uniformly within the range $(0,1]$.
The final time $T$ was chosen to be sufficiently large to achieve
full controllability over $\U(N)$ at $t=T$ \cite{Raj2007,Ramakrishna1995}. The observed trends are consistent
with theoretical predictions since $m_1 = 2N-2 < 2N-1$, whereas $m_2= 4N-4 \gg 2N-1$.

An increase was observed in the mean condition numbers
encountered for large $m$ MOTC algorithms starting from the field
modes of $\varepsilon(t)$ tuned to the transition frequencies of the
system, compared to $\varepsilon(t)$ modes tuned to sample
frequencies within roughly the same range but not precisely
tuned to the transition frequencies (data not shown; fields of the
form (\ref{randfield}) with frequencies $\omega_j =
0.1j({E_N-E_1})/\hbar,\quad 1 \leq j \leq 50$ were used). Thus
careful tuning of the initial guess for the control may facilitate
implementation of MOTC algorithms for large $m$.

\subsection{MOTC with multiple measurement bases}

\begin{figure}
\centerline{
\includegraphics[width=5.5in,height=4in]{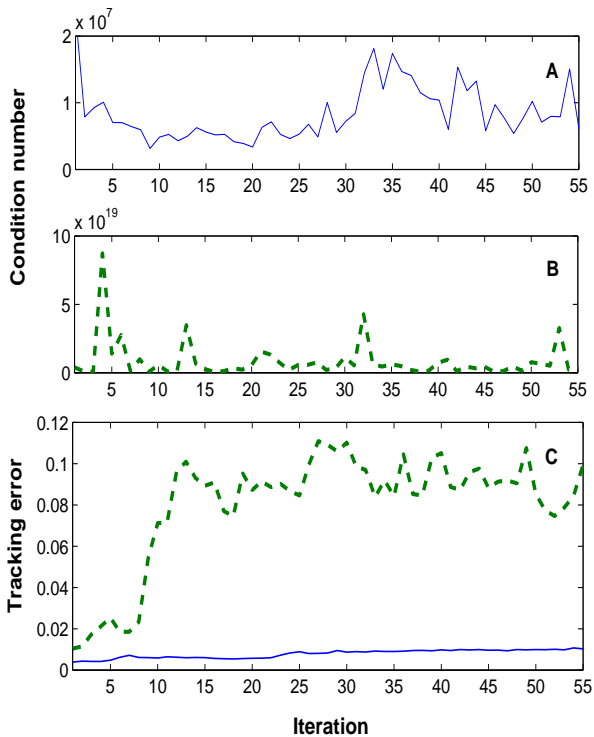}
}
\caption{Comparison of MOTC of a set of 40 observables from four measurement
bases for a full-rank nondegenerate mixed initial state $\rho(0)$ and for a pure initial
state; the lowest eigenstate $E_1$, $|\psi_{1}\rangle$ of $H_0$).
\textbf{A)} $\Gamma$ matrix condition numbers for thermal $\rho(0)$; \textbf{B)} $\Gamma$ matrix condition
numbers for pure $\rho(0)$;  \textbf{C)} Associated tracking errors (solid = thermal, dashed = pure). 
}\label{meas40}
\end{figure}

\begin{figure}
\centerline{
\includegraphics[width=4in,height=4in]{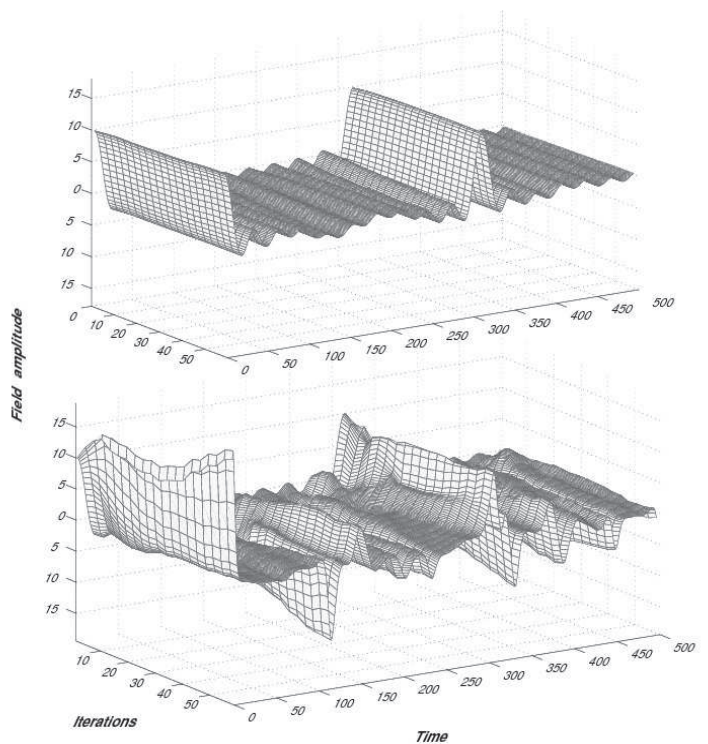}
} \caption{Evolution of the control field during MOTC with different
numbers of measurement bases (system (\ref{system example})). \textbf{Top:} 5 observables
(1 measurement basis); \textbf{Bottom:} 15 observables (2 measurement bases).
$\varepsilon_0(t)$ was of the form (\ref{randfield}); $\rho(0)$ was
a full-rank, nondegenerate mixed state.
}\label{fieldevol}
\end{figure}

\begin{figure}
\centerline{
\includegraphics[width=4in,height=2.8in]{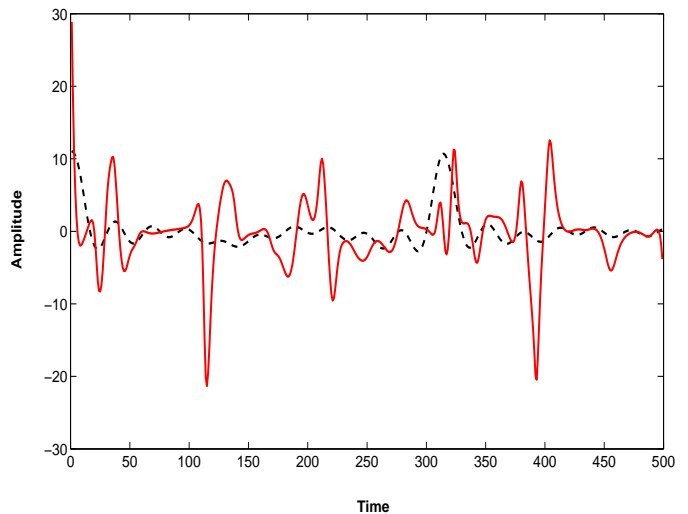}
}
\caption{Superimposition of the optimal control fields identified by MOTC
for $5$ (dashed) and $20$ (solid) observable control problems.
$\varepsilon_0(t)$ was of the form (\ref{randfield}); $\rho(0)$ was
a full-rank, nondegenerate mixed state. Observables were drawn successively
from mutually unbiased measurement bases as described in equation (\ref{mub}).
}\label{fieldcomp5-20}
\end{figure}

\begin{figure}
\centerline{
\includegraphics[width=4.7in,height=3.4in]{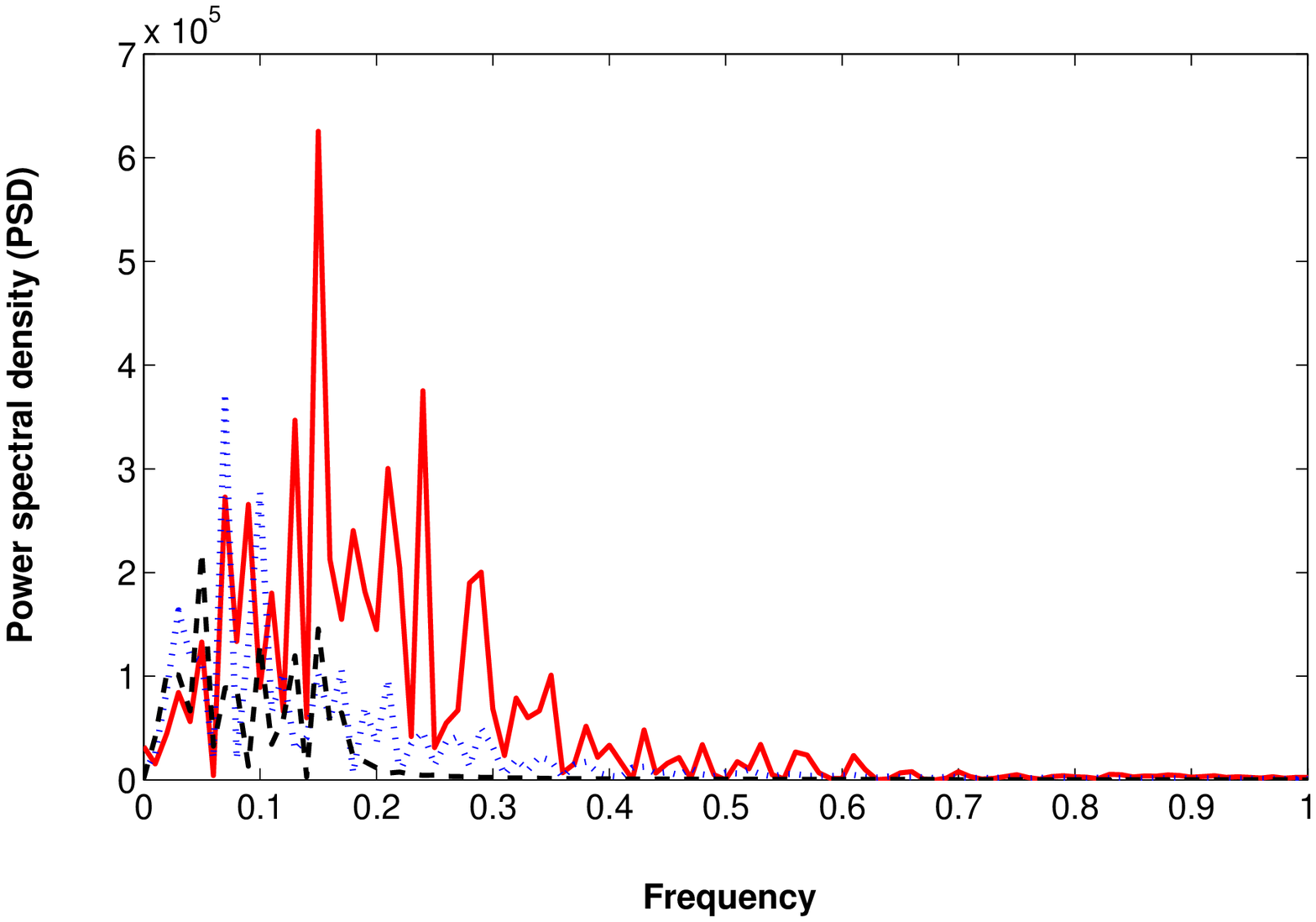}
}
\caption{Fourier power spectra of optimal control fields obtained by MOTC tracking for
various numbers of observables $m$.
The spectra for $m=5$ (dashed) and $m=20$ (solid),
corresponding to the fields in Fig. \ref{fieldcomp5-20},
are shown alongside the spectrum for $m=15$ (dotted).}\label{ft_5-15-20}
\end{figure}

Greater numbers of orthogonal observables can be effectively tracked
for state preparation $\rho(0)$ of higher rank. A full set of
observables from multiple measurement bases (or more generally, $m \leq
2Nn-n^2$ orthogonal observables) can be effectively tracked for
state preparation of $\rho(0)$ with rank $n$, according to equation (\ref{max m and rank rho}).

Figs. \ref{gramfig2}A and B indicate that for $m=40$, the $\Gamma$ condition
numbers for thermal $\rho(0)$ are significantly lower than those for
pure $\rho(0)$.  It should be possible to track 40 observables
without significant numerical errors occurring due to singularity of the MOTC Gramian. 40 observables can be
tracked through measurements in four bases. For this purpose, we
employed the $r=0,1,2,3$ MUB bases given by equation (\ref{mub}).
The $\Gamma$ condition numbers and associated tracking errors are
presented in Fig. \ref{meas40}A,C (solid trace). Despite uniformly
higher condition numbers, singularity is not a limiting factor when $\rho(0)$ is thermal; the
error correction methods described in section \ref{expt}
suffice to maintain the desired optimization trajectory.

For $\rho(0)$ being a pure state, the number of observables that can be
simultaneously tracked along arbitrary paths while avoiding
$\Gamma$-matrix singularities is lowest. According to Fig.
\ref{gramfig2}, the $\Gamma$-matrix for tracking 40 observables for $\rho(0)$ being pure should be effectively
singular. This leads to unacceptably large tracking errors and deviation from the expected
path, as shown in Fig. \ref{meas40}C. Indeed, based on the analytical relation ({\ref{max m and rank rho}),
it should not be possible to track a full set of observables from more than two measurement bases when
$\rho(0)$ is pure. In these cases, search algorithms based on scalar objective optimization, such as the gradient, may
be preferred. Nonetheless, as described in Section III, gradient search will generally not effectively locate points on the Pareto front for arbitrary sets of multiple observables.

The evolution of control fields along the optimization trajectory
for increasing numbers of observables sheds light on the impact of
multiple observable objectives on the control landscape. Fig.
\ref{fieldevol} compares the control field evolutions for $m=5$ and $m=15$ MOTC,
with the track $\textbf{w}_s$ given by (\ref{geotrack}) between $U_0$ and a target
propagator lying on the global maximum manifold $\mathcal{M}_1^{\max}$ for the first observable.
At each step, the two fields reside on the same level set of the first observable, but generally
deviate increasingly from each other on the domain $\U(N)$. The
imposition of additional observable constraints delimits each
successive level set such that it contains a smaller set of more
complex fields. Fig. \ref{fieldcomp5-20} isolates and compares the optimal fields obtained by
$m=5$ and $m=20$ MOTC.  The power spectra of the optimal control fields for $m=5$, $m=15$ and $m=20$ are shown
in Fig. \ref{ft_5-15-20}. Note that although
these fields lie on the same level set of a single observable, they are more diverse than those in Fig. \ref{paretofield}, which lie on distinct submanifolds of the Pareto front $\mathcal{P}_w^{\e}$. Since control of more orthogonal observables requires control over more independent parameters of the unitary propagator $U(T)$, the latter problem imposes more constraints on the permissible dynamics over $0 \leq t \leq T$.

Note that all the MOTC variants in this section (control of up to 40 observables) can be
implemented through measurements of four or fewer full-rank, nondegenerate observables in mutually unbiased measurement
bases.

\section{Discussion}\label{discussion}

To date, the majority of quantum optimal control experiments (OCE)
have been implemented with adaptive (e.g., genetic) search algorithms, which require only measurement of the expectation value of an observable at the final dynamical time for each successive control field. Recently,
gradient flow algorithms have been implemented in OCE studies, based on the finding that quantum control
landscapes are devoid of suboptimal traps \cite{Roslund2008}, and the observation in computer simulations
that gradient-based algorithms converge more efficiently than genetic algorithms.
Similar conclusions may be drawn about search algorithms for quantum multiobservable
control. Whereas in generic multiobjective optimization problems, the existence of local traps
typically requires the use of stochastic methods, the inherent monotonicity of quantum control
landscapes enables more efficient deterministic algorithms for multiobjective quantum control.
Quantum multiobservable tracking control (MOTC) is an example of such an algorithm.

Efficient control algorithms are especially important in the context of quantum Pareto optimization, since the tradeoff between observable objectives in such control problems renders the problem of identifying
optimal control fields on the Pareto front via stochastic algorithms much more expensive than
the analogous problem of single observable control. Moreover, in quantum Pareto optimization, it is generally
nontrivial to parameterize a family of scalar objective functions whose maxima are all (weak) Pareto optima, as
demonstrated in section \ref{kinpareto}. Promising successes have been reported in the application of multiobjective
evolutionary algorithms (MOEA), which avoid this obstacle, to quantum Pareto optimal control problems involving a
limited number of quantum observables $m$ \cite{Wolf2007}. However, the convergence time of such algorithms scales
unfavorably with system size due to the computational expense of sorting nondominated solutions \cite{Deb2002}. By
contrast, apart from the overhead costs of initial state estimation and online measurements of observable expectation
values, the convergence time of MOTC does not scale with the number of observables optimized.

In quantum multiobservable maximization problems, it is particularly efficient to first identify points on
the kinematic Pareto front $\mathcal{P}_w^{U}$ of the multiobservable optimization problem, which requires
knowledge or estimation of the initial density matrix $\rho(0)$ of the quantum
system. In the case that $\rho(0)$ is not a thermal mixed state, it can be determined by the method of maximal likelihood estimation (MLE) of quantum states \cite{Banaszek1999}. Once $\rho(0)$ is known, the Pareto optima on $\U(N)$ can be sampled efficiently using the methods described in Section
\ref{kinpareto}. Observable tracking can then be applied in open loop quantum control experiments
to locate optimal control fields $\e^*(t)$ on the dynamic Pareto front $\mathcal{P}_w^{\e}$.
The local vicinity of each $\e^*(t)$ can subsequently be explored to identify controls with the
most favorable dynamic properties.

Experimental observable expectation value tracking requires an extension of gradient flow methods that have
already been applied in the laboratory \cite{Roslund2008}. Specifically, instead of following the path of steepest ascent, the control field is updated in a direction that in the first-order approximation would produce the next observable track value. In principle, the ability to follow the path of steepest ascent for single observable maximization should be extensible to following arbitrary trajectories across multiobservable control landscapes. When applied in conjunction with signal averaging, statistical estimation of the gradient - employing as few 30 sampled points as per iteration - has been found to confer adequate accuracy and robustness to laser noise in single observable gradient flow algorithms. In practice, experimental multiobservable tracking may demand additional robustness to noise, due to the smaller number of control fields that are consistent with any given point along the track. Experimental error correction, described in Section IV and applied numerically in Section VI, may therefore be essential for these applications. Our goal in this paper has been to lay the theoretical foundations for quantum Pareto optimal control and to illustrate its implementation under ideal conditions. Numerical studies that simulate the effects of control field noise \cite{Shuang2004} on MOTC will be reported in a separate work.

The eigenvalue spectrum of the initial density matrix $\rho(0)$ affects the combinations of observable expectation values that can be simultaneously achieved and the nonsingularity of the multiobservable control Gramian. Tracking a vector $\vec \Phi(s)$ of $m$ orthogonal observable expectation values rarely encounters singularities originating in lack of local controllability, if $m \ll N^2-1$. However, for the greater number of observables that must be controlled in complex Pareto optimization or state preparation problems, the MOTC Gramian is more prone to becoming ill-conditioned. This effect can be alleviated by carefully tuning the initial guess for the control field.

The additional overhead required for multiobservable tracking, relative to stochastic
algorithms, can be mitigated by the development of efficient estimators of multiobservable
control gradients. We have presented simple measurement techniques, including the use of mutually unbiased measurement
bases, that can dramatically improve MOTC efficiency for large numbers of observables. Still, the methodology for
gradient estimation described in section
\ref{orthog} does not exploit the statistical correlation of components of the multiobservable
gradient (\ref{gradmult}), which originates in the correlation between elements of $\rho(0)$.
Future work should investigate the application of MLE to generate more accurate estimates of the
multiobservable gradient for an equivalent number of observable measurements.
In addition, it is important to examine the impact of quantum decoherence - which produces nonunitary
dynamics described by Kraus operators - on the efficiency of multiobservable tracking.

\bigskip

\appendix

\section{Characterization of Pareto optimal submanifolds}

In order to qualitatively assess the likelihood of the gradient flow (\ref{Eflow}) converging to a weak Pareto optimum where the expectation value of observable $k$ is maximized, the dimensions of  $\mathcal{I}_{k,M}^{\max,\max} = \mathcal{M}_k^{\max}\bigcap \mathcal{M}_M^{\max} $ and $\mathcal{M}_M^{\max}$ may be compared. If the two dimensions are equal, there is a finite chance of the gradient flow converging to a Pareto optimum. Otherwise,  although the gradient flow may converge arbitrarily close to a Pareto optimum, the attractor of the flow will not be in $\mathcal{P}_w^{(U,\e)}$ and the likelihood of approaching within a fixed radius of a Pareto optimal point will generally be diminished. In this Appendix we present a few results regarding the intersections of the critical submanifolds of the observable objective functions (\ref{obs}), to facilitate such an analysis.
As a special case, we also derive the conditions guaranteeing convergence to a weak Pareto optimum given in Lemma 1.

Denote by $\tilde \Theta_a$ the diagonal form of $\Theta_a$ in a
basis $S_a$ where its eigenvalues are arranged in increasing order;
i.e., $\tilde \Theta_a\equiv S_a^{\dag}\Theta_a S_a=
diag\{\gamma_{a(1)},\cdots,\gamma_{a(1)};\cdots;\gamma_{a(s_a)},\cdots,\gamma_{a(s_a)}\}$,
according to the convention in Section III. Set the diagonal matrix
$\tilde \Theta_{b} \equiv S_a^{\dag} \Theta_{b}
S_a=S_a^{\dag}S_b\cdot
diag\{\gamma_{b(1)},\cdots,\gamma_{b(1)};\cdots;\gamma_{b(s_{b})},\cdots,\gamma_{b(s_{b})}\}\cdot
S_b^{\dag} S_a.$ Let $\U(\textbf{m}_b') = S_a^{\dag}S_b
\U(\textbf{m}_{b}) S_b^{\dag} S_a$, where $\U(\textbf{m}_{b})$ is
defined as in Section III.

We first obtain an upper bound on $\mathrm{dim~}\mathcal{I}_{a,b}^{ij}$.
For any observable $\Theta_a$, it was shown in \cite{WuMike2008} that the dimension of
the critical manifold $\mathcal{M}_{a}^{i}$ of $\Phi_a$ is given by
\begin{equation}
\mathrm{dim~}\mathcal{M}_{a}^{i} = \sum_{x=1}^r n_x^2 + \sum_{y=1}^{s_{a}} m_{ay}^2 - \sum_{x=1}^r \sum_{y=1}^{s_{a}}v_{xy}^2,
\end{equation}
where the $v_{xy}$ are overlap numbers that denote the number of
positions where the eigenvalues $\gamma_{ax}$ and $\gamma_{ay}$
appear simultaneously in $\tilde \rho(0)$ and $\tilde \Theta_a$, and
the $n_x$, $m_{ky}$ are defined as in Section III. Then, we have
$\mathrm{dim~}\mathcal{I}_{a,b}^{ij} \leq {\min}
\left\{\mathrm{dim~}\mathcal{M}_{a}^{i},\mathrm{dim~}\mathcal{M}_{b}^{j}\right\}$
for the upper bound.

Further analytical characterization of $\mathcal{I}_{a,b}^{ij}=
\mathcal{M}_a^{i} \bigcap \mathcal{M}_{b}^{j}$, including a lower
bound on its dimension, is possible when the two observables
$\Theta_a$ and $\Theta_{b}$ commute. According to equation
(\ref{quotient}) we have $\mathcal{M}_a^{i} \bigcap
\mathcal{M}_b^{j} = \U(\textbf{n})\pi^i \U(\textbf{m}_a) \bigcap
\U(\textbf{n})\pi^j \U(\textbf{m}_b')$ for any $\pi^i \in
\mathcal{D}_a^i$, $\pi^j \in \mathcal{D}_{b}^j$.  This can be
expressed
\begin{widetext}
\begin{equation*}
\left\{\bigcup_{\pi \in \mathcal{D}_{\cap}^{ij} } 
\U(\textbf{n})\pi \left(\U(\textbf{m}_a) \cap \U(\textbf{m}_b')\right)\right\} ~~\bigcup ~~\left\{\bigcup_{\pi^i \not \in \mathcal{D}_{\cap}^{ij}~ \lor~ \pi^j \not \in \mathcal{D}_{\cap}^{ij}} \U(\textbf{n})\pi^i \U(\textbf{m}_a) ~\cap~ \U(\textbf{n})\pi^j \U(\textbf{m}_b')\right\},
\end{equation*}
\end{widetext}
where $\mathcal{D}_{\cap}^{ij}\equiv \mathcal{D}_a^i \cap
\mathcal{D}_b^j$. The first term in this expression can  be written
$\{U\pi V ~ |~ U \in \U(\textbf{n});~V \in  \U(\textbf{m}_a) \cap
\U(\textbf{m}_b');~\pi \in \mathcal{D}_{\cap}^{ij} \}$. Denote the
Lie subgroup $\U(\textbf{m}_a) ~\cap~ \U(\textbf{m}_b')$ of $\U(N)$
by $\U(\textbf{P})$. Then, we have
\begin{equation}\label{intersection}
\bigcup_{\pi \in \mathcal{D}_{\cap}^{ij} } \U(\textbf{n})  ~\pi ~\U(\textbf{P})~ \subseteq ~ \mathcal{I}_{a,b}^{ij}.
\end{equation}
Let us partition the set $\mathcal{D}_{\cap}^{ij}$ with the
equivalence relation $\sim$ defined by $\pi_1 \sim \pi_2$ if $\pi^2
= U \pi^1 V,$ for $(U,V) \in  \U(\mathbf{n}) \times \U(\mathbf{P})$.
Then it is straightforward to show that $\pi$'s in different
equivalence classes with respect to $\sim$ label disconnected
submanifolds of  $\bigcup_{\pi \in \mathcal{D}_{\cap}^{ij} }
\U(\textbf{n})  \pi \U(\textbf{P})$. $\U(\textbf{P})$ can be written
$\U(p_{11})\times\cdots \times \U(p_{s_as_b})$, where the $p_{xy}$
are overlap numbers that denote the number of positions where the
eigenvalues $\gamma_{ax}$ and $\gamma_{by}$  appear simultaneously
in $\tilde \Theta_a$ and $\tilde \Theta_b'$. Together, these overlap
numbers form a contingency table \cite{WuMike2008}, with rows
indexed by $x$ and columns indexed by $y$. The row sums are
$\sum_{y=1}^{s_b}p_{xy} = m_{ax}$ and the column sums are
$\sum_{x=1}^{s_a}p_{xy}=m_{by}$.

Based on these results we can now derive conditions guaranteeing convergence of (\ref{Eflow}) to a weak Pareto optimum. We have $\mathcal{M}_b^{j} \subseteq \mathcal{M}_a^{i}$ if there is just one nonzero entry in each column of the contingency table (such that each $S_a^{\dag}S_b ~\U(m_{by}) ~S_b^{\dag} S_a \subseteq \U(m_{ax})$ for just one $x$, hence $\U(\textbf{P}) = \U(\textbf{m}_b')$), and $\mathcal{D}_b^j \subseteq \mathcal{D}_a^i$ (such that $\mathcal{D}_a^i \bigcap \mathcal{D}_b^j = \mathcal{D}_b^j$); then, $\mathcal{I}_{a,b}^{ij} = \mathcal{M}_{b}^{j}$.
In this case we have the first equality in equation (\ref{intersection}) and all $\pi \in \mathcal{D}_{\cap}^{ij}$ belong to the same equivalence class with respect to $\sim$. If we set $\Theta_a = \Theta_k$ and $\Theta_b = \Theta_M$, this proves the conditions guaranteeing convergence to $\mathcal{P}_w^{(U,\e)}$ in Lemma 1. $\blacksquare$

We now obtain a lower bound on $\mathrm{dim~}\mathcal{I}_{a,b}^{ij}$
in the case of commuting $\Theta_a$, $\Theta_b$. Define
$F_{\pi}(P,Q): \pi \rightarrow P \pi Q,$ where $(P,Q) \in
\U(\textbf{n}) \times \U(\textbf{P})$, and let $stab~(F_{\pi})$
denote the set of all $(U,V) \in \U(\textbf{n}) \times
\U(\textbf{P})$ such that $F_{\pi}(U,V) = U\pi V = \pi$. It is
straightforward to verify \cite{WuMike2008} that
$stab~(F_{\pi})=\U(\textbf{n}) ~\bigcap~ \pi^{\dag}~ \U(\textbf{P})
~\pi$. The latter may be written $\U(\textbf{Q})=\U(q_{111})\times
\cdots \times \U(q_{rs_as_b})$, where the overlap numbers $q_{xyz}$
denote the number of positions where the eigenvalue $\lambda_{i}$
appears simultaneously with $\gamma_{ay}$ and $\gamma_{bz}$, after
the permutation $\pi$ has been applied to $\tilde \Theta_a,~\tilde
\Theta_{b}'$. It was shown in \cite{WuMike2008} that any manifold of
the type $\U(\textbf{n})  ~\pi ~\U(\textbf{P})$ has dimension
$\mathrm{dim~}\U(\textbf{n})+\mathrm{dim~}\U(\textbf{P})-\mathrm{dim~}
stab(F_{\pi})$. Given that the dimension of $\U(N)$ is $N^2$, the
lower bound on the dimension of the intersection submanifold is thus
\begin{widetext}
\begin{equation}\label{interdim}
\mathrm{dim~}\mathcal{I}_{a,b}^{ij} \geq \max_{\pi \in \mathcal{D}_{\cap}^{ij} }\left\{
\sum_{x=1}^r n_x^2 + \sum_{x=1}^{s_a}\sum_{y=1}^{s_{b}} p_{xy}^2 - \sum_{x=1}^r \sum_{y=1}^{s_a}\sum_{z=1}^{s_{b}}q_{xyz}^2\right\}.
\end{equation}
\end{widetext}
In the case that the observables commute, we therefore have both analytical lower and upper
bounds on $\mathrm{dim~}\mathcal{I}_{k,M}^{\max,\max}$, whereas if
they do not commute, we have only the upper bound. If the upper
bound is less than $\mathrm{dim~}\mathcal{M}_M^{\max}$, then the
chance of the flow converging to $\mathcal{P}_w^{(U,\e)}$ is
infinitesimally small (although it may approach arbitrarily
close). The lower bound is most useful when it equals
$\mathrm{dim~}\mathcal{M}_M^{\max}$, in which case there is a finite
chance that the flow will converge to $\mathcal{P}_w^{(U,\e)}$.


\end{document}